\documentclass[AMA,Times1COL]{WileyNJDv5} 
\usepackage{bm}

\articletype{Preprint}%

\received{Date Month Year}
\revised{Date Month Year}
\accepted{Date Month Year}
\journal{Preprint}
\volume{00}
\copyyear{2026}
\startpage{1}

\definecolor{Maroon}{RGB}{128,0,0}
\definecolor{Orange}{RGB}{255,165,0}
\definecolor{Green}{RGB}{0,128,0}
\definecolor{Blue}{RGB}{0,0,255}

\begin{document}

\title{Matrix-free phase-field modeling of fracture in micromechanical testing simulations of inelastic materials}

\author[1]{Fabio Di Gioacchino}

\author[1]{Rezgar Shakeri}

\author[1]{Zachary R. Atkins}

\author[1]{Karen Stengel}

\author[1]{Layla Ghaffari}

\author[1]{Jeremy L. Thompson}

\author[1]{Jed Brown}

\authormark{DI GIOACCHINO \textsc{et al.}}
\titlemark{Matrix-free phase-field modeling of fracture for micromechanical testing simulations of inelastic materials}

\address[1]{\orgdiv{Computer Science Department}, \orgname{University of Colorado at Boulder}, \orgaddress{\state{Colorado}, \country{USA}}}

\corres{Fabio Di Gioacchino, 
\email{fabio.digioacchino@colorado.edu}}



\abstract[Abstract]{Resolving steep damage gradients across diffuse cracks in the phase-field modeling of fracture favors the use of high-order finite elements, for which matrix-free methods can provide superior performance and scalability. Here, we implement the Perić \& Dettmer constitutive framework for visco-elastoplastic materials in an open source solid mechanics library supporting matrix-free operators for high-order finite elements with $p$-multigrid preconditioning on GPUs. We introduce a rheological \textit{fracture element} assembled in series with the Perić \& Dettmer block, keeping fracture and inelastic material properties mutually independent. We conduct numerical simulations of tensile and compressive quasistatic tests on synthetic particle-matrix microstructures using an El Capitan high-performance computing prototype. The results reproduce representative inelastic responses and fracture patterns, including X-shaped cracking observed in polymer-bonded explosives and cup-and-cone fracture in ductile metals.}


\jnlcitation{\cname{%
\author{Di Gioacchino F},
\author{Shakeri R},
\author{Atkins ZR},
\author{Stengel K},
\author{Ghaffari L}, 
\author{Thompson JR}, and
\author{Brown J}}.
\ctitle{Matrix-free phase-field modeling of fracture in micromechanical testing simulations of inelastic materials} \cjournal{\it J XX.} \cvol{202X;00(00):X--XX}.}

\maketitle


\renewcommand\thefootnote{\fnsymbol{footnote}}
\setcounter{footnote}{1}

\section{Introduction}\label{sec1}

Traditional approaches to modeling fracture in finite element simulations treat cracks as discontinuities in the displacement field, requiring explicit tracking of crack surfaces and \textit{ad hoc} criteria for crack propagation \cite{belytschko_review_2009}. Phase-field fracture modeling eliminates these requirements by introducing a continuous scalar field that regularizes the sharp discontinuity over a finite length. The diffused crack topology enables a variational formulation of Griffith's fracture theory \cite{francfort_variational_2008} that is amenable to both standard and emerging finite element methods \cite{kakouris_phase-field_2017, aldakheel_phase-field_2018, hirshikesh_adaptive_2019, li_extension_2025}. 


Early developments focused on brittle fracture in elastic solids to address crack irreversibility and propagation under compressive loads, distinguishing models by their dissipation structure and whether an elastic threshold precedes damage initiation \cite{ambati_review_2015}. These studies linked the regularization length scale to measurable quantities, such as characteristic lengths, highlighting analogies with gradient damage theories \cite{de_borst_gradient_2016, tanne_crack_2018}. Due to ease of implementation and capacity to capture complex crack branching and merging patterns, phase-field formulations of brittle fracture have since been implemented in open source software \cite{hirshikesh_fenics_2019,  kumar_gpfnics_2023, sidharth_open-source_2024, castillon_phasefieldx_2025, barki_phafidyn_2025, munshi_detailed_2026} and extended to cohesive and dynamic fracture \cite{borden2012dynamic, ren_explicit_2019, mandal2020review_dynamic_pf, feng_phase-field_2022}, anisotropic crack propagation \cite{nguyen_multi-phase-field_2017, bleyer_phase-field_2018, li_crack_2019, liu2025anisotropic_silicon_pf, ruan2025anisotropic_pf_am}, cyclic loading (fatigue) \cite{lo_phase-field_2019, carrara_framework_2020, yang_acceleration_2023, li_review_2023, liu_crystal_2025}, and multiphysics problems \cite{cui2022generalised_multiphasefield_scc, wang2023explicit_hydrogen_pf, zhang2024multiphysics_mg_scc}. However, materials used in structural applications exhibit microstructural features designed to maximize strength while enabling dissipative mechanisms that delay damage accumulation, such as plastic yielding. This raises fundamental questions regarding the energetic and dissipative interplay between inelastic deformation and damage evolution in these materials.

\begin{figure*}[!h]%
  \centerline{\includegraphics[width=300pt,height=14pc,trim=5 5 5 5,clip]{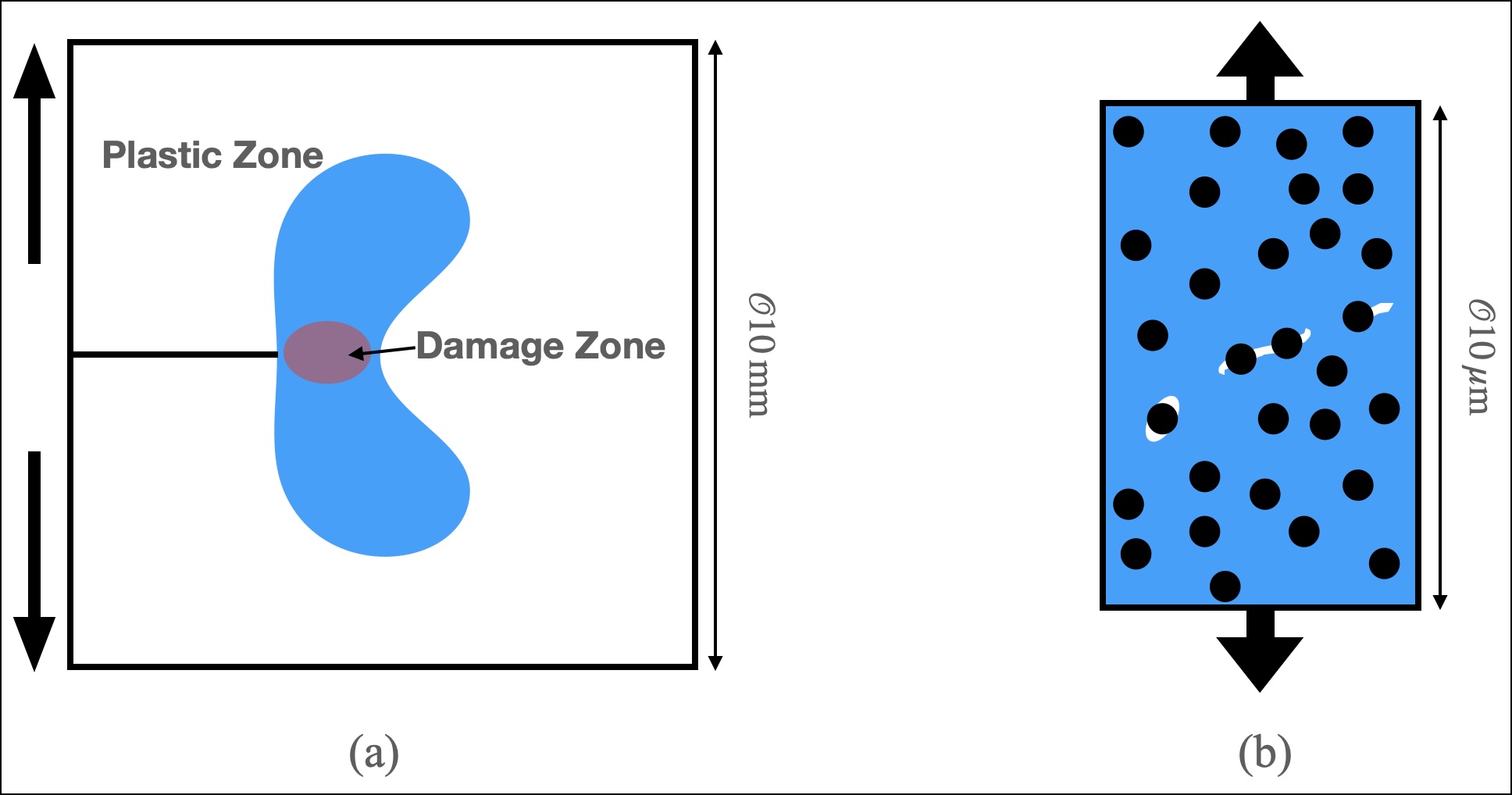}}
  \caption{Schematic distinction between plastic and damage zones at different length scales. (a) At homogenization length scales, the damage zone represents the region containing microvoids and microcracks. (b) At the microstructural length scale, the damage zone coincides with the location of microvoids and microcracks within the microstructure.}
  \label{fig_schematic}
  \end{figure*}

For a homogenized material volume, such as the notched tensile specimen of an inelastic material illustrated in Fig. \ref{fig_schematic}a, regions of inelastic deformation and damage spatially overlap. Following Lemaitre's continuum damage mechanics \cite{lemaitre_continuous_1985} and the Gurson–Tvergaard–Needleman micromechanical theories \cite{gurson_continuum_1977,tvergaard_analysis_1984}, a damage variable  is introduced that can be interpreted as a measure of the effective reduction in load-bearing cross-sectional area resulting from the nucleation and coalescence of microvoids and microcracks that are not spatially resolved. Accordingly, damage-induced degradation can extend beyond elastic properties to include yield strength, strain hardening behavior, and viscosity parameters, for example, through the introduction of a plastic degradation function \cite{ambati2015phase, talamini_attaining_2021, svolos_phase-field_2025}. Conversely, inelastic deformation can influence damage evolution through local stress concentrations arising from inelastic strain incompatibilities between microstructural constituents, such as grains in polycrystals \cite{di_gioacchino_experimental_2015, edwards_experimental_2019} or precipitates \cite{jones_reduced_2018}. This, in turn, has motivated the development of constitutive models that incorporate inelastic energy contributions into the damage driving force \cite{miehe2016phase, han2022crack, abrari_vajari_micropolar_2025} or allow fracture toughness to decrease with accumulated inelastic deformation \cite{han2022crack, yin2020ductile, khalil2022generalised}. At the microstructural length scale of investigation, on the other hand, damage zones can be directly identified with regions where microvoids nucleate and microcracks propagate, as illustrated in Fig. \ref{fig_schematic}b. This motivates a formulation in which fracture and inelastic material properties are prescribed independently, \textit{i.e.}, without explicitly linking fracture toughness to accumulated inelastic strain or inelastic material parameters to damage. A corresponding rheological representation would assign fracture to a dedicated element assembled in series with the inelastic constitutive block.

Despite straightforward physical interpretation and improved predictive capability, studies on phase-field fracture modeling in inelastic materials with resolved microstructures are largely limited to 2D representations. Full 3D applications \cite{zhang_3d_2020, cheng_wavelet-enriched_2020, rezwan2025_coupled_pf_cp, xiao2025_phasefield_fracture_SiC_Al} remain challenging due to the high computational cost associated with the fine mesh resolutions required to capture microstructural features and the steep damage gradients that arise across cracks. When linear shape functions are used, a sufficient number of finite elements of size $h$ must span half the crack width $l$, \textit{i.e.} $h \ll l$. By contrast, the convergence study by Jodlbauer \textit{et al.} \cite{jodlbauer_matrix-free_2020} demonstrated that high-order discretizations can provide an accurate solution with $h \approx l$, and without the spurious oscillations observed for linear elements. As shown in the same study, the computational demands of high-order finite elements motivate adopting matrix-free methods, which evaluate operator actions through element-level computations without explicitly assembling or storing global sparse operator matrices. These methods can increase arithmetic intensity and reduce memory traffic relative to assembled approaches, enabling more effective utilization of GPUs. Combining matrix-free representations with multigrid preconditioners has demonstrated near-optimal scaling in (hyper)elasticity \cite{Davydov2020MatrixFreeHyperelastic, brown_performance_2022, Schussnig2024MatrixFreeHP} and phase-field modeling of brittle fracture \cite{jodlbauer_parallel_2020, jodlbauer_matrix-free_2020, liu_efficient_2022, kolditz_matrix-free_2025}, Table \ref{table_literature}.

\begin{figure*}[!h]%
  \centerline{\includegraphics[width=210pt,height=12pc, trim=8 8 8 8,clip]{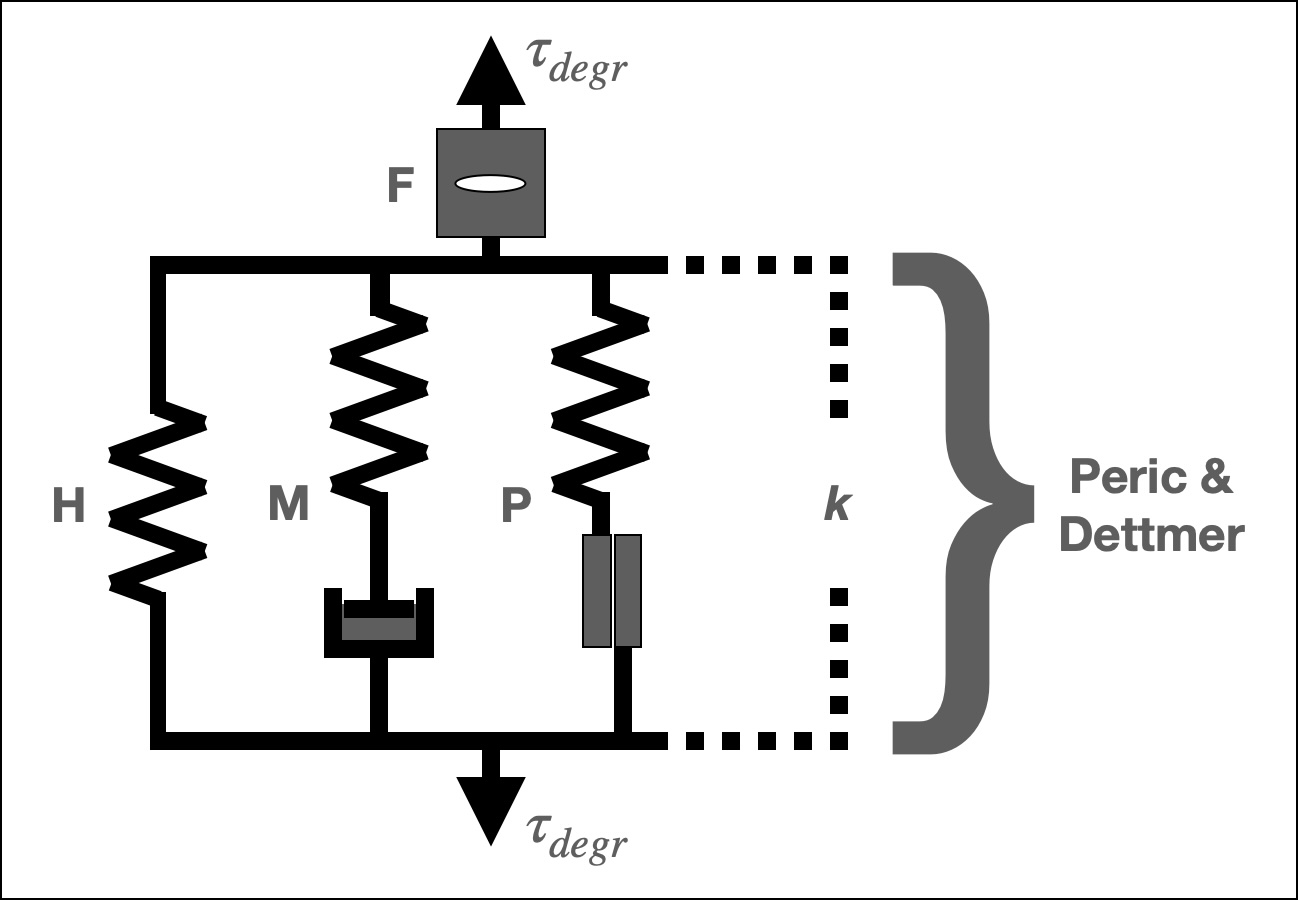}}
  \caption{Assembly of rheological elements for inelasticity and fracture mechanics. In the Perić \& Dettmer block, multiple Hooke (H), Maxwell (M), and Prandtl (P) branches can be assembled in parallel. A rheological fracture element (F), depicted as a solid containing a sharp crack oriented perpendicularly to the uniaxial stress direction, is added in series.}
  \label{fig_rheological_model}
  \end{figure*}

The contribution of the present work is twofold: 
(i) We combine phase-field modeling of fracture with the Perić \& Dettmer \cite{peric_computational_2003} framework, which constructs inelastic constitutive models of isotropic materials through the parallel assembly of Hooke (elastic), Maxwell (viscoelastic) and Prandtl (elasto-plastic) rheological branches. In particular, we introduce a rheological \textit{fracture element} that is assembled in series, as illustrated in Fig. \ref{fig_rheological_model}. Damage accumulation deactivates the Perić \& Dettmer block without affecting the underlying inelastic material properties at the dashpot and slider. Similarly, it is the stress associated with the purely elastic strain energy stored in the block that drives damage. Together, these aspects realize the constitutive response for the microstructural length scale scenario depicted in Fig. \ref{fig_schematic}b above.
(ii) We implement the proposed framework in the \texttt{Ratel} open source solid mechanics library \cite{atkins_ratel_2026}, which is built on top of libCEED \cite{Brown2021} and PETSc \cite{petsc-user-ref} to natively support matrix-free operators and data structures optimized for GPU-accelerated high performance computing (HPC) platforms. The solver employs $p$-multigrid preconditioning, which has proven robust for finite element discretizations on unstructured meshes and integrates naturally with matrix-free methods for computational efficiency and parallel scalability \cite{brown_performance_2022}. As shown in Table \ref{table_literature}, the present work expands current efforts in matrix-free solid mechanics to both inelasticity and phase-field modeling of fracture with GPU use. 

We report results of proof-of-concept numerical experiments on synthetic particle-matrix microstructures of compressible materials, in which characteristic inelastic responses and crack patterns are qualitatively reproduced. Numerical simulations are carried out on an El Capitan HPC prototype, demonstrating the suitability of \texttt{Ratel} for solving large scale solid mechanics problems on next-generation supercomputing platforms. We conclude with a summary of the main features and findings and discuss future extensions to anisotropic behavior in polycrystals, with applications to micromechanical testing of interest.

  \begin{table*}[!h]%
    \centering %
    \caption{Present studies on the phase-field modeling of fracture with matrix-free methods.\label{table_literature}}%

    \begin{tabular*}{\textwidth}{@{\extracolsep\fill}lllllllll@{\extracolsep\fill}}
    \toprule
    \textbf{Ref.} & \textbf{year} & \textbf{monolithic scheme} & \textbf{finite strains} & \textbf{inelasticity} & \textbf{iterative solver}  & \textbf{preconditioning} & \textbf{mesh adaptivity} & \textbf{GPUs} \\
    \midrule
   Jodlbauer \textit{et al.} \cite{jodlbauer_parallel_2020} & 2020 & $\checkmark$  & $\times$   & $\times$  & GMRES & AMG and GMG & $\times$  & $\times$  \\
   Jodlbauer \textit{et al.} \cite{jodlbauer_matrix-free_2020} & 2020  & $\checkmark$  & $\times$   & $\times$  & GMRES & AMG and GMG & $\times$  & $\times$   \\
   Liu \textit{et al.} \cite{liu_efficient_2022} & 2022 & $\times$   & $\times$   & $\times$   &CG & AMG & $\times$  & $\times$   \\
   Koldvitz \textit{et al.} \cite{kolditz_matrix-free_2025} & 2025 & $\checkmark$  & $\times$  & $\times$  & GMRES & GMG & $\checkmark$ & $\times$   \\
   \textit{Present Study} & -  & $\checkmark$   & $\checkmark$  & $\checkmark$ & GMRES & $p$-multigrid and AMG & $\times$   & $\checkmark$ \\ 
    \bottomrule
    \end{tabular*}

    \end{table*}  

\section{Peric \& Dettmer rheology for Hencky materials}\label{sec2}

We specialize the Perić \& Dettmer multiplicative plasticity and viscoelasticity framework for Hencky materials. As shown in Perić \textit{et al.} \cite{peric_model_1992}, the multiplicative basis provides a physical response under non-coaxial loads, unlike large strain additive formulations. By adopting Hencky strain energy functionals and an exponential map for time integration, the stress updating procedure in current configuration conveniently reuses the classical return mapping algorithm for small strains, with additional pre- and post-processing steps to map strains and stresses between Euclidean and logarithmic strain spaces. Furthermore, by restricting updates to deviatoric components in a principal stress-based format, the framework ensures an efficient implementation of models that can capture a wide range of isotropic mechanical responses with a minimal set of material parameters.

\subsection{Preliminaries}
Supported by experimental observations of plastic deformation in crystalline materials, the deformation gradient tensor $\bm{F} = \bm{I} + \nabla_X \bm{u}$ 
can be multiplicatively decomposed into inelastic $\bm{F}^{in}$ and elastic $\bm{F}^e$ parts such that \cite{Lee1969}
\begin{equation}
\bm{F} = \bm{F}^e \bm{F}^{in}
\label{eq:inelasticity-def-grad-multiplicative-decomposition}
\end{equation}

Viscoelastic ($\bm{F}^v = \bm{F}^{in}$) or plastic ($\bm{F}^p = \bm{F}^{in}$) deformations are assumed to be isochoric (\textit{i.e.}, $\det \bm{F}^{in} = 1$), mapping material points to an intermediate configuration. 
Hence, $\bm{F}^e = \textbf{v}^e \bm{R}^e$ maps the intermediate configuration to the spatial configuration through the elastic stretch $\textbf{v}^e$ and underlying rotation $\bm{R}^e$. 
With Eq. \eqref{eq:inelasticity-def-grad-multiplicative-decomposition}, the spatial velocity gradient $\bm{l} \equiv \frac{\partial \bm{v}}{\partial \bm{x}} = \dot{\bm{F}} \bm{F}^{-1}$ can be additively decomposed as
\begin{equation}
  \bm{l} = \bm{l}^e + \bm{F}^e \bm{L}^{in} (\bm{F}^e)^{-1}
  \label{eq:inelasticity-velocity-gradient}
\end{equation}
where $\bm{l}^e \equiv \dot{\bm{F}}^e (\bm{F}^e)^{-1}$ and $\bm{L}^{in} \equiv \dot{\bm{F}}^{in} (\bm{F}^{in})^{-1}$, which lives in the intermediate configuration. The Lie derivative of the left elastic Cauchy-Green tensor $\bm{b}^e = \bm{F}^e \bm{F}^{eT}$ with respect to the velocity field $\bm{v}$ can thus be expressed as
\begin{equation}
  \mathcal{L}_{\bm{v}} \bm{b}^e 
  := \bm{F} \frac{\partial}{\partial t} 
  \left( \bm{F}^{-1} \bm{b}^e \bm{F}^{-T} \right) 
  \bm{F}^T 
  = \dot{\bm{b}}^e - \bm{l} \bm{b}^e - \bm{b}^e \bm{l}^T
  \label{eq:inelasticity-Lie-derivative-be}
  \end{equation}
  
Eq. \eqref{eq:inelasticity-Lie-derivative-be} helps derive a constitutive framework from the (isothermal) Clausius-Duhem inequality
  \begin{equation}
    \bm{\tau}: \bm{d} - \dot{\psi} \geq 0
  \label{eq:inelasticity-clausius-duhem}
  \end{equation} 
where $\bm{\tau}$ is the Kirchhoff stress work conjugate of the rate of deformation tensor $\bm{d} \equiv \mathrm{sym}(\bm{l})$ and $\psi(\bm{b}^e,  \bm{\alpha})$ is the Helmholtz free energy density, which depends on elastic strains and a set of internal variables $\bm{\alpha}$ associated with dissipative mechanisms. Following derivations in Simo and Hughes \cite{simo1998computational}, the time derivative of $\psi$ can be obtained using Eq. \eqref{eq:inelasticity-Lie-derivative-be} as
\begin{align}
    \dot{\psi}(\bm{b}^e, \bm{\alpha})
    = \frac{\partial \psi}{\partial \bm{b}^e} : \dot{\bm{b}}^e 
       + \frac{\partial \psi}{\partial \bm{\alpha}} \cdot \dot{\bm{\alpha}}
    = 2 \frac{\partial \psi}{\partial \bm{b}^e} \bm{b}^e 
       : \left( \bm{d} 
       + \tfrac{1}{2} (\mathcal{L}_{\bm{v}} \bm{b}^e) 
         \bm{b}^{e-1} \right)
       + \frac{\partial \psi}{\partial \bm{\alpha}} \cdot \dot{\bm{\alpha}}
    \label{eq:inelasticity-free-energy-rate}
    \end{align}
Substituting Eq. \eqref{eq:inelasticity-free-energy-rate} into Eq. \eqref{eq:inelasticity-clausius-duhem} and rearranging terms yields
    \begin{equation}
    \left( \bm{\tau} - 2 \frac{\partial \psi}{\partial \bm{b}^e} 
    \bm{b}^e \right) : \bm{d} 
    + 2 \frac{\partial \psi}{\partial \bm{b}^e} \bm{b}^e 
      : \left( -\tfrac{1}{2} (\mathcal{L}_{\bm{v}} \bm{b}^e) 
      \bm{b}^{e-1} \right)
    - \frac{\partial \psi}{\partial \bm{\alpha}} \cdot \dot{\bm{\alpha}}
    \ge 0
    \label{eq:inelasticity-clausius-duhem-expanded}
    \end{equation}
Standard arguments then lead to the constitutive equation for $\bm{\tau}$ and a simplified dissipation inequality,
    \begin{equation}
    \bm{\tau} = 2 \frac{\partial \psi}{\partial \bm{b}^e} 
    \bm{b}^e, 
    \qquad
    \bm{\tau} : 
    \left( -\tfrac{1}{2} (\mathcal{L}_{\bm{v}} \bm{b}^e) 
    \bm{b}^{e-1} \right)
    - \frac{\partial \psi}{\partial \bm{\alpha}} \cdot \dot{\bm{\alpha}}
    \ge 0
    \label{eq:inelasticity-clausius-duhem-results}
    \end{equation}
An isotropic, fourth order and positive definite tensor $\mathsf{V}$ can finally be introduced to trivially satisfy Eq. \eqref{eq:inelasticity-clausius-duhem-results} in the absence of internal variables \cite{ReeseGovindjee1998FiniteViscoelasticity},
    \begin{equation}
    -\tfrac{1}{2} (\mathcal{L}_{\bm{v}} \bm{b}^e) 
    \bm{b}^{e-1} 
    := \mathsf{V} : \bm{\tau}
    \label{eq:inelasticity-evolution-law}
    \end{equation}

\subsection{Exponential Map Integrator}

At a given global Newton iterate, the current deformation gradient $\bm{F}_{n+1}$ is held fixed during the local constitutive corrector. Consequently, the velocity gradient associated with the corrector is taken to vanish ($\bm{l}=\bm{0}$), and Eq. \eqref{eq:inelasticity-Lie-derivative-be} reduces to $\mathcal{L}_{\bm{v}}\bm{b}^e=\dot{\bm{b}}^e$. Substituting into Eq. \eqref{eq:inelasticity-evolution-law} and rearranging yields
    \begin{equation}
    \dot{\bm{b}}^e = -2  (\mathsf{V} : \bm{\tau})  \bm{b}^e
    \label{eq:inelasticity-be-rate-constitutive}
    \end{equation}
which can be solved using a backward Euler approximation based on the exponential mapping scheme \cite{eterovic1990,weber1990},
    \begin{equation}
    \bm{b}^e = \exp \left( -2  \Delta t  \mathsf{V} : \bm{\tau} \right) \bm{b}^{e \, \mathrm{tr}}
    \label{eq:inelasticity-exponential-mapping}
    \end{equation}
The tensor exponential maps traceless tensors $\bm{X}$ to unimodular tensors because $\det(\exp(\bm{X})) = \exp [\mathrm{tr}(\bm{X})]= 1$. It therefore preserves the isochoric character of (deviatoric) plastic deformation, unlike alternative integration schemes, such as the one adopted in the phase-field modeling of ductile fracture in Borden \textit{et al.}  \cite{borden2016phase}.

In Eq \eqref{eq:inelasticity-exponential-mapping}, the superscript “tr” denotes trial quantities associated with the generally inadmissible (purely) elastic increment as the initial guess. The trial elastic left Cauchy strain tensor $\bm{b}^{e \, \mathrm{tr}}$ is computed after retrieving the inverse of the inelastic right Cauchy strain tensor (with $\bm{C} = \bm{F}^T \bm{F}$), from the previously converged time step, $\bm{C}^{in-1}_n$, as
    \begin{equation}
    \bm{b}^{e \, \mathrm{tr}} = \bm{F}  \bm{C}^{in-1}_n  \bm{F}^{T}
    \label{eq:inelasticity-be-update}
    \end{equation}
With $\bm{b}^e$ in Eq. \eqref{eq:inelasticity-exponential-mapping}, the corresponding logarithmic strain is computed as
\begin{equation}
  \bm{\varepsilon}^e
  = \frac{1}{2} \sum_{i=1}^{3} \log(b^e_i)  \hat{\bm{n}}_i \otimes \hat{\bm{n}}_i
  \label{eq:inelasticity-log-strain}
\end{equation}
where $b^e_i$ are the eigenvalues of $\bm{b}^e$, which are related to the principal stretches as $b^e_i = \lambda_i^{e \ 2}$, and $\hat{\bm{n}}_i$ are the shared eigenvectors of $\bm{b}^e$, $\bm{\varepsilon}^e$, and $\tau$, as follows from the coaxiality of \textit{elastic} strain and stress tensors in isotropic constitutive models. To ensure numerical stability when $\lambda^e_i \approx 1$, we implement  Eq. \eqref{eq:inelasticity-log-strain} using the \texttt{log1p} function. As shown in Shakeri \textit{et al.} \cite{shakeri2024stable}, this adds the intermediate step of computing Hencky strains using the eigenvalues of the Green-Lagrange or Euler-Almansi strain tensors.

In addition to attractive physical properties, logarithmic strain measures cancel out the exponential of an isotropic tensor function in Eq. \eqref{eq:inelasticity-exponential-mapping}, leading to a small strain format for the update of strain components. As we limit inelasticity to deviatoric components $\bm{\varepsilon}_\mathrm{d}^e$, this gives
    \begin{equation}
    \bm{\varepsilon}_\mathrm{d}^e = \bm{\varepsilon}_\mathrm{d}^{e \, \mathrm{tr}} - \Delta t  (\mathsf{V} : \bm{\tau}_\mathrm{d})
    \label{eq:inelasticity-deviatoric-strain-update}
    \end{equation}
where $\bm{\tau}_\mathrm{d} = \bm{\tau} - \frac{1}{3} \mathrm{tr}{(\bm{\tau}})\bm{I}$ is the deviatoric stress.
As inelastic contributions are isochoric, the volumetric strain $\bm{\varepsilon}_\mathrm{v} = \bm{\varepsilon}_\mathrm{v}^\mathrm{e}$ and $ \mathrm{tr}(\bm{\varepsilon}^{\mathrm{e} \, \mathrm{tr}}) = \mathrm{tr}(\bm{\varepsilon}^e)$ ($= \mathrm{tr}(\bm{\varepsilon})$). The elastic strain is then obtained as
    \begin{equation}
\bm{\varepsilon}^e = \bm{\varepsilon}_\mathrm{d}^{e} + \bm{\varepsilon}_\mathrm{v} = \bm{\varepsilon}_\mathrm{d}^{e} + \frac{\mathrm{tr}(\bm{\varepsilon}^{\mathrm{e} \, \mathrm{tr}})}{3}\bm{I}
    \label{eq:inelasticity-deviatoric-strain-update-2}
  \end{equation}
With $\bm{b}^e = \exp(2 \bm{\varepsilon}^e)$, the state variable is finally updated as
    \begin{equation}
\bm{C}^{in-1}_{n+1} = \bm{F}^{-1}  \bm{b}^e  \bm{F}^{-T}
    \label{eq:inelasticity-deviatoric-strain-update-3}
  \end{equation}
and stored to be used in the subsequent time step.

\subsection{Parallel Assembly of the Perić \& Dettmer Rheological Block}
    
\begin{figure*}[h]%
    \centering
    \includegraphics[width=250pt,height=15pc, trim=8 8 10 10,clip]{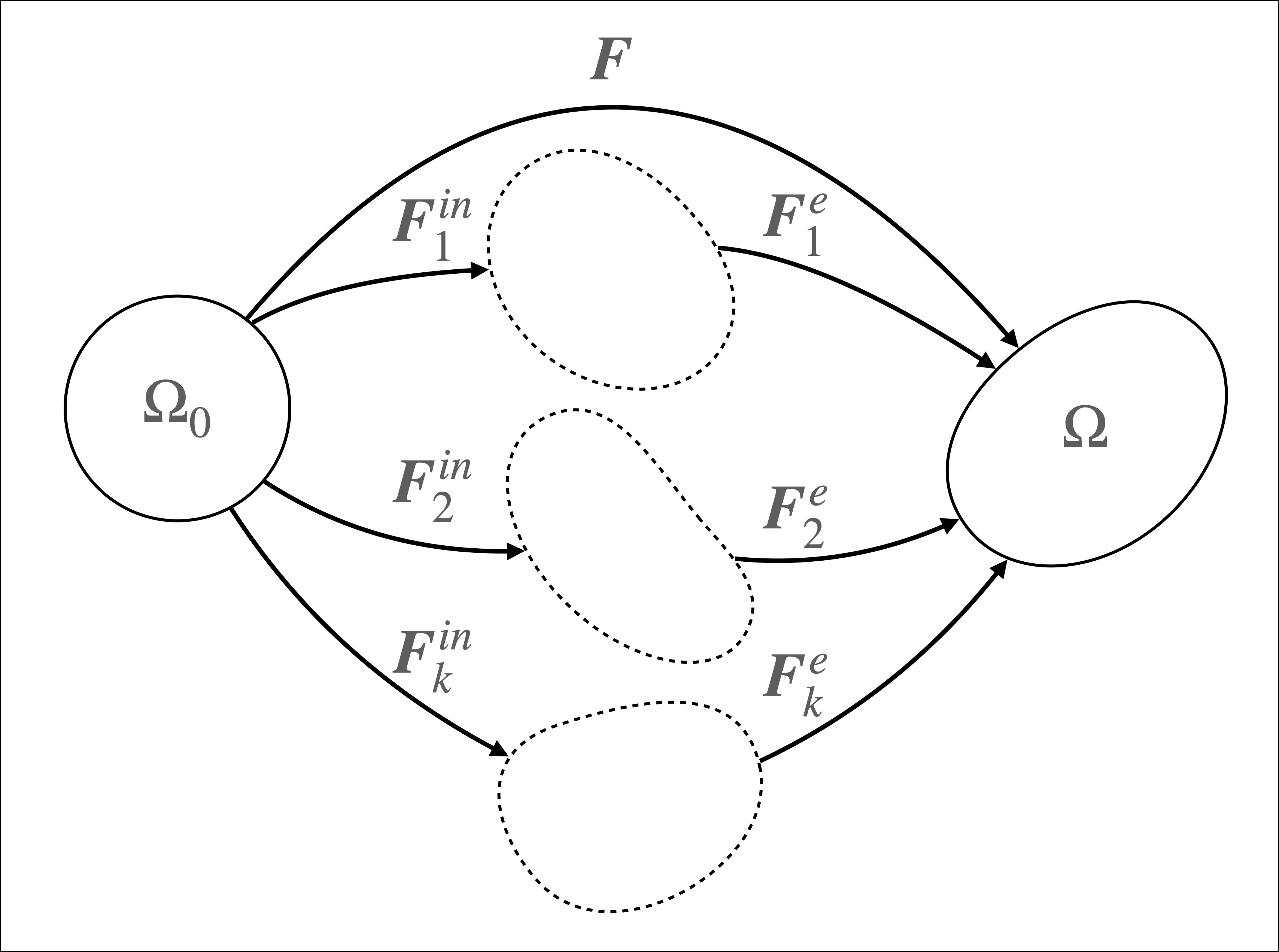}
    \caption{Multiplicative decomposition of the total deformation gradient for the Perić \& Dettmer rheological block.\label{fig_def_grad_tens}}
\end{figure*}

The above results can be generalized to $k$ rheological components arranged in parallel for which  $\bm{F} = \bm{F}^e_k \bm{F}^{in}_k$, as illustrated in Fig. \ref{fig_def_grad_tens}. The total free energy can thus be expressed as the sum of the free energies of individual components, $\psi = \sum_k \psi_k (\bm{b}^e_k, \bm{\alpha}_k)$, and the (total) Kirchhoff stress as $\bm{\tau} = \sum_k \bm{\tau}_k$ with
\begin{equation}
    \bm{\tau}_k = 2 \frac{\partial \psi_k}{\partial \bm{b}^e_k}  \bm{b}^e_k,
    \qquad
    \mathrm{and}
    \qquad
    \bm{\tau}_k :
    \mathsf{V}_k : \bm{\tau}_k
    - \frac{\partial \psi_k}{\partial \bm{\alpha}_k} \cdot \dot{\bm{\alpha}}_k \ge 0
    \label{eq:inelasticity-assembly-rheology}
\end{equation}

Evolution laws for plastic deformation and internal variables specific to each rheological element are derived from thermodynamic arguments to ensure thermodynamic consistency. Within the numerical treatment, these laws result in algorithmic expressions for the stress update that depend on trial quantities
\begin{equation}
\bm{\tau}_k = \bm{\tau}_k (\bm{\alpha}_n, \, \bm{\varepsilon}^{e \, \mathrm{tr}}_{n+1})
\label{eq:inelasticity-consistent-stress}
\end{equation}
The algorithmic update is linearized to obtain a consistent tangent, which enters the Jacobian to preserve the quadratic convergence of Newton’s method \cite{simo1998computational}:
    \begin{equation}
    \frac{\partial \bm{\tau}_k}{\partial \bm{F}_{n+1}}
    = \frac{\partial \bm{\tau}_k}{\partial \bm{\varepsilon}^{e}_{n+1}} : 
    \underline{\frac{\partial\bm{\varepsilon}^{e}_{n+1}}{\partial \bm{\varepsilon}^{e \, \mathrm{tr}}_{n+1}} }: 
    \frac{\partial\bm{\varepsilon}^{e \, \mathrm{tr}}_{n+1}}{\partial \bm{b}^{e \, \mathrm{tr}}_{n+1}} : 
    \frac{\partial \bm{b}^{e \, \mathrm{tr}}_{n+1}}{\partial \bm{F}_{n+1}}
    \label{eq:inelasticity-consistent-tangent}
    \end{equation}
where the underscored term depends on the specific rheological branch. 

Boxes 1 and 2 summarize the stress update algorithm and its derivation, which is presented below in full tensor form. By taking advantage of the coaxiality between the strain and stress tensors in isotropic constitutive models, the update can be formulated with respect to the principal directions. This representation significantly reduces the number of floating point operations, as most of the tensorial updates reduce to scalar operations on eigenvalues. 
Equations in the principal stresses-based format implemented here are reported in Appendix A. The linearized terms, also derived in principal directions, are provided in Appendix B.

\begin{boxwithhead}
  {BOX 1\quad General stress update algorithm for $k$-rheological branch}
  {\noindent 
  \begin{enumerate}
  \item For each branch, retrieve $\bm{C}^{in -1}_n$ from the previously converged time step. With current deformation gradient $\bm{F}$, compute the trial deviatoric part of the logarithmic strain:
  \begin{equation}
  \begin{aligned}
  \bm{\varepsilon}^{e \, \mathrm{tr}}_\mathrm{d} &= \bm{\varepsilon}^{e \, \mathrm{tr}} - \frac{\mathrm{tr}(\bm{\varepsilon}^{e \, \mathrm{tr}})}{3}\bm{I} 
  \qquad \mathrm{with} \qquad
  \bm{\varepsilon}^{e \, \mathrm{tr}} &= \tfrac{1}{2} \log(\bm{b}^{e \, \mathrm{tr}})
  \qquad \mathrm{and} \qquad
  \bm{b}^{e \, \mathrm{tr}} &= \bm{F}  \bm{C}^{in-1}_n  \bm{F}^{T}
  \end{aligned}
  \label{eq:plasticity-elastic-predictor}
  \end{equation}
  
  \item Correction step:

    \textbf{Hooke branch:}
    $$
    \bm{\varepsilon}_\mathrm{d}^e=\bm{\varepsilon}_\mathrm{d}^{e \, \mathrm{tr}}
    $$

    \textbf{Maxwell branch:}
    $$
      \bm{\varepsilon}_\mathrm{d}^e = \left( 1 + \mu\frac{\Delta t}{\eta_\mathrm{d}} \right)^{-1} 
      \bm{\varepsilon}_\mathrm{d}^{e \, \mathrm{tr}}
    $$

    \textbf{Prandtl branch:} Compute $q^{\mathrm{tr}}=\sqrt{3 J_2(\bm{\tau}_\mathrm{d}^{\mathrm{tr}})}$ and retrieve $\bar{\varepsilon}^p_n$ to evaluate $\Delta\gamma$ (Box 2). Substitute in
    $$
    \bm{\varepsilon}_\mathrm{d}^e = \left(1 - 3\mu \frac{\Delta\gamma}{q^{\mathrm{tr}}}\right)\bm{\varepsilon}_\mathrm{d}^{e \, \mathrm{tr}}
    $$
    $$
    \bar{\varepsilon}^p_{n+1} = \bar{\varepsilon}^p_n + \Delta\gamma
    $$

  \item Updated Kirchhoff stress as
  $$
  \bm{\tau} = 2 \mu  \bm{\varepsilon}_\mathrm{d}^e + \kappa  \mathrm{tr}(\bm{\varepsilon}^e)  \bm{I}
  $$

  \item Update $\bm{C}^{in-1}$ and store for subsequent time step:
  $$
  \bm{C}^{in-1}_{n+1} 
  = \bm{F}^{-1}  \bm{b}^e  \bm{F}^{-T}
  \qquad \mathrm{with} \qquad
  \bm{b}^e = \exp(2 \bm{\varepsilon}^e)
  $$
\end{enumerate}
}
  \end{boxwithhead}

\begin{boxwithhead}
  {BOX 2\quad Prandtl branch $\Delta \gamma$ evaluation}
  {\noindent
  \begin{enumerate}
    \item   With current $q^{\mathrm{tr}}$ and $\bar{\varepsilon}^p_n$, check stress state admissibility:
    \begin{equation}
      \begin{aligned}
      &\mathrm{if} \quad q^{\mathrm{tr}} - \sigma_y(\bar{\varepsilon}^p_n) \le 0 : \quad \texttt{GO TO 2} \\
      &\mathrm{else:} \quad \texttt{GO TO 3}
      \end{aligned}
      \label{eq:plasticity-admissibility-check}
      \end{equation}
    
    \item \textbf{Elastic step}. Set $\Delta \gamma = 0$.
    
    \item \textbf{Plastic step}. Solve
    \begin{equation}
      \begin{aligned}
    \tilde{f} 
    = q^{\mathrm{tr}} - 3\mu \Delta\gamma 
    - \sigma_y(\bar{\varepsilon}^p_n + \Delta\gamma) = 0
    \label{eq:plasticity-yield-function-trial-state-1}
    \end{aligned}
    \end{equation}
    for $\Delta\gamma$ via Newton-Raphson iterations to
    $|\tilde{f}| \le \mathrm{tol}$:
    \begin{equation}
    \begin{aligned}
    H &:= \frac{d\sigma_y}{d\bar{\varepsilon}^p}\Big|_{\bar{\varepsilon}^p_n+\Delta\gamma} \\
    den &:=\frac{d\tilde{f}}{d\Delta\gamma} = -3\mu - H \\
    \Delta\gamma &\leftarrow \Delta\gamma - \frac{\tilde{f}}{den}
    \end{aligned}
    \label{eq:plasticity-delta-gamma-update}
    \end{equation}
  \end{enumerate} 
}
  \end{boxwithhead}
  
\subsubsection{Hooke Branch}

Hencky hyperelasticity adopts the small strains quadratic form of the strain energy density $\psi$ with respect to logarithmic strains, which can also be expressed as a deviatoric $\psi_\mathrm{d}$ and volumetric  $\psi_\mathrm{v}$ energy split,
\begin{equation}
\psi(\bm{\varepsilon^e}) = \tfrac{1}{2}  \bm{\varepsilon}^e : \mathsf{C} : \bm{\varepsilon}^e
= \psi_\mathrm{d} + \psi_\mathrm{v} =\mu \bm{\varepsilon}_\mathrm{d}^e : \bm{\varepsilon}_\mathrm{d}^e +\frac{\kappa}{2}  [\mathrm{tr}(\bm{\varepsilon}^e)]^2
\label{eq:hencky-free-energy}
\end{equation}
where $\mathsf{C}$ is the fourth order elasticity tensor, and $\mu$ and $\kappa$ are the shear and bulk moduli, respectively. The expression for the Kirchhoff stress preserves the small strain format:
\begin{equation}
\bm{\tau} = 2 \frac{\partial \psi}{\partial \bm{b}^e}  \bm{b}^e
= 2 \frac{\partial \psi}{\partial \bm{\varepsilon}^e}
  \frac{\partial \bm{\varepsilon}^e}{\partial \bm{b}^e} \bm{b}^e
= 2 \frac{\partial \psi}{\partial \bm{\varepsilon}^e} \frac{1}{2}
\frac{\partial \log \bm{b}^e}{\partial \bm{b}^e}  \bm{b}^e
= \frac{\partial \psi}{\partial \bm{\varepsilon}} = 2 \mu  \bm{\varepsilon}_\mathrm{d}^e + \kappa  \mathrm{tr}(\bm{\varepsilon}^e)  \bm{I}
\label{eq:hencky-tau-current}
\end{equation}
For a Hooke branch (\textit{i.e.}, a single spring), there are no inelastic contributions and $ \bm{\varepsilon}^e=\bm{\varepsilon}$.

\subsubsection{Maxwell Branch}

As the amount of viscoelastic stress relaxation depends on the current state, a thermodynamically consistent implementation of a Maxwell branch can be obtained without prescribing internal variables (\textit{i.e.}, $\bm{\alpha}$ = $\bm{0}$) for Eq. \eqref{eq:inelasticity-clausius-duhem-results}, 
and

\begin{equation}
\mathsf{V} : \bm{\tau}_\mathrm{d} = \frac{1}{2 \eta_\mathrm{d}}  \bm{\tau}_\mathrm{d}
\label{eq:viscoelasticity-evolution-law}
\end{equation}
where $\eta_\mathrm{d}$ $[Pa \cdot s]$ is the deviatoric viscosity\cite{ReeseGovindjee1998FiniteViscoelasticity}. 
Equation \eqref{eq:inelasticity-deviatoric-strain-update} becomes
\begin{equation}
\bm{\varepsilon}_\mathrm{d}^e = \bm{\varepsilon}_\mathrm{d}^{e \, \mathrm{tr}} - \frac{\Delta t}{2 \eta_\mathrm{d}}  \bm{\tau}_\mathrm{d}
\label{eq:viscoelasticity-deviatoric-strain-update}
\end{equation}
As $\bm{\tau}_\mathrm{d}$ is generally a nonlinear function of $\bm{\varepsilon}_\mathrm{d}^e$, Eq. \eqref{eq:viscoelasticity-deviatoric-strain-update} is commonly solved via local Newton iterations. Nevertheless, it reduces to a closed form update for Hencky materials \cite{fancello2006variational},
\begin{equation}
\bm{\varepsilon}_\mathrm{d}^e = \left( 1 + \mu\frac{\Delta t}{\eta_\mathrm{d}} \right)^{-1} 
\bm{\varepsilon}_\mathrm{d}^{e \, \mathrm{tr}}
\label{eq:viscoelasticity-deviatoric-strain-update-hencky}
\end{equation}

\subsubsection{Prandtl Branch}

To formulate constitutive laws governing the evolution of internal variables and plastic flow that satisfy the dissipation inequality, plasticity relies on a dissipation potential $\Xi$ that is convex in both stresses and internal variables \cite{simo1998computational}. In associative plasticity, the dissipation potential coincides with the yield function, $f$. The principle of maximum plastic dissipation then implies that the plastic flow vector 
$\bm{N}^p$ is normal to the yield surface $f = 0$,
\begin{equation}
\bm{N}^p :=
\frac{\partial \Xi}{\partial \bm{\tau}} 
= \frac{\partial f}{\partial \bm{\tau}}
\label{eq:plasticity-flow-vector}
\end{equation}
Thermodynamic consistency in Eq. \eqref{eq:inelasticity-clausius-duhem-results} is maintained by setting
\begin{equation}
  \mathsf{V} : \bm{\tau}_\mathrm{d}
= \dot{\gamma} \frac{\partial f}{\partial \bm{\tau}}
= \dot{\gamma} \bm{N}^p,
\qquad
\dot{\bm{\alpha}} = 
\dot{\gamma} \frac{\partial f}{\partial \bm{A}}
\label{eq:plasticity-flow-rule}
\end{equation}
where $\dot{\gamma}$ is the plastic multiplier determined through the standard optimality (Kuhn-Tucker) conditions,
\begin{equation}
\dot{\gamma} \ge 0, 
\qquad 
f \le 0, 
\qquad 
\dot{\gamma} f = 0
\label{eq:plasticity-optimality-conditions}
\end{equation}
and $\bm{A}$ denotes the set of forces conjugate to the internal variables,
\begin{equation}
\bm{A} = - \frac{\partial \psi}{\partial \bm{\alpha}}
\label{eq:plasticity-work-conjugate}
\end{equation}

The definition of $f$ and its associated return mapping algorithm depend on the particular choice of the yield criterion. Motivated by experimental observations of pressure insensitive yielding in metals, plastic deformation in von Mises plasticity occurs when the deviatoric strain energy exceeds a critical value. This energy can be expressed in terms of the second deviatoric stress invariant
\begin{equation}
J_2 = \tfrac{1}{2} \bm{\tau}_\mathrm{d} : \bm{\tau}_\mathrm{d} = 2\mu{\psi_\mathrm{d}}
\label{eq:plasticity-J2}
\end{equation}
The critical value is given by the strain hardening law 
$\sigma_y = \sigma_y(\bar{\varepsilon}^p_n)$, which is a function of the accumulated (effective or equivalent) plastic strain $\bar{\varepsilon}^p$, defined as
\begin{equation}
\bar{\varepsilon}^p = \int_0^t |\dot{\gamma}| dt
\label{eq:plasticity-accumulated-plastic-strain}
\end{equation}
Since uniaxial tensile testing is typically used to calibrate plasticity models, the von Mises yield function is usually expressed in terms of the yield stress $\sigma_y$ as
\begin{equation}
f = q - \sigma_y
\label{eq:plasticity-vonmises-yield}
\end{equation}
where the scalar quantity 
$q = \sqrt{3J_2(\bm{\tau}_\mathrm{d})} = \sqrt{\tfrac{3}{2}} \|\bm{\tau}_\mathrm{d}\|$ 
is the von Mises (effective or equivalent) stress. The implemented strain hardening law combines linear and Voce nonlinear hardening as
\begin{equation}
\sigma_y = \sigma_0 
+ H_{lin} \bar{\varepsilon}^p 
+ (\sigma_\infty - \sigma_0) \big(1 - \mathrm{exp}({-\beta \bar{\varepsilon}^p})\big)
\label{eq:plasticity-strain-hardening-law}
\end{equation}
where $\sigma_0$ is the initial yield stress and 
$\{H_{lin}, \sigma_\infty, \beta\}$ are the three hardening parameters, respectively representing the linear hardening modulus, the saturation (flow) stress, and the hardening decay parameter. Eq. \eqref{eq:plasticity-strain-hardening-law} is evaluated using the accumulated plastic strain from the previously converged time step $\bar{\varepsilon}^p_n$, which is updated as,
\begin{equation}
  \bar{\varepsilon}^p_{n+1} = \bar{\varepsilon}^p_n + \Delta\gamma
  \label{eq:plasticity-accumulated-plastic-update}
  \end{equation}

If plastic deformation occurs, \textit{i.e.} $f > 0$, the flow vector takes the form
\begin{equation}
\bm{N}^p
= \sqrt{\tfrac{3}{2}} \frac{\bm{\tau}_\mathrm{d}}{\|\bm{\tau}_\mathrm{d}\|}
= \tfrac{3}{2} \frac{\bm{\tau}_\mathrm{d}}{q}
\label{eq:plasticity-flow-vector-vonmises}
\end{equation}
Substituting Eq. \eqref{eq:plasticity-flow-vector-vonmises} into Eq. \eqref{eq:plasticity-flow-rule} yields the expression for the update of the deviatoric strain components,
\begin{equation}
  \begin{aligned}
  \bm{\varepsilon}_\mathrm{d}^e
  = \bm{\varepsilon}_\mathrm{d}^{e \, \mathrm{tr}} 
     - \Delta t\left( \frac{\Delta\gamma}{\Delta t} \bm{N}^{p \ \mathrm{tr}} \right)
  = \bm{\varepsilon}_\mathrm{d}^{e \, \mathrm{tr}} 
     - \Delta\gamma\left(\tfrac{3}{2} \frac{\bm{\tau}_\mathrm{d}^{\mathrm{tr}}}{q^{\mathrm{tr}}}\right) 
  = \left(1 - 3\mu \frac{\Delta\gamma}{q^{\mathrm{tr}}}\right)
     \bm{\varepsilon}_\mathrm{d}^{e \, \mathrm{tr}}
  \end{aligned}
  \label{eq:plasticity-deviatoric-strain-update}
  \end{equation}
Combining the updates in Eq. \eqref{eq:plasticity-accumulated-plastic-update} and Eq. \eqref{eq:plasticity-deviatoric-strain-update} with Eq. \eqref{eq:plasticity-vonmises-yield}, the return mapping equations for von Mises plasticity reduce to a single nonlinear equation \cite{simo1998computational}:
\begin{equation}
\tilde{f} 
= q^{\mathrm{tr}} - 3\mu \Delta\gamma 
- \sigma_y(\bar{\varepsilon}^p_n + \Delta\gamma) = 0
\label{eq:plasticity-yield-function-trial-state-2}
\end{equation}
which can be efficiently solved for $\Delta\gamma$ by Newton-Raphson iterations in Box 2.


\subsubsection{Perfect Plasticity}

In the parallel rheological assembly, the linear hardening contribution to strain hardening ($H_{lin}$ in Eq. \eqref{eq:plasticity-strain-hardening-law}) can be assigned to the Hooke (spring) branch. Similarly, in isotropic materials with spatially resolved microstructures, nonlinearities in the overall mechanical response generally arise from microstructural features that promote heterogeneous plastic yielding across the material volume. Viscoelastic effects and damage accumulation further accentuate this nonlinear response. Therefore, a rheological assembly that includes perfectly plastic Prandtl branch(es) can capture nonlinear strain hardening with fewer material parameters.

In the perfect plastic regime, the consistency condition yields an explicit expression for $\Delta\gamma$,
\begin{equation}
  \Delta\gamma
  = \frac{q^{\mathrm{tr}} - \sigma_0}{3\mu}
  \label{eq:perfect-plastic-delta-gamma}
\end{equation}
Substituting Eq. \eqref{eq:perfect-plastic-delta-gamma} into the deviatoric strain update in Eq. \eqref{eq:plasticity-deviatoric-strain-update} gives the closed form for perfect plasticity,
\begin{equation}
  \begin{aligned}
  \bm{\varepsilon}_\mathrm{d}^e 
  = \left(1 - 3\mu\frac{\Delta\gamma}{q^{\mathrm{tr}}}\right)
     \bm{\varepsilon}_\mathrm{d}^{e \, \mathrm{tr}}
  = \left(1 - \frac{q^{\mathrm{tr}} - \sigma_0}{q^{\mathrm{tr}}}\right)
     \bm{\varepsilon}_\mathrm{d}^{e\,\mathrm{tr}}
  = \left(\frac{\sigma_0}{q^{\mathrm{tr}}}\right)
     \bm{\varepsilon}_\mathrm{d}^{e\,\mathrm{tr}}
  \end{aligned}
  \label{eq:plasticity-deviatoric-strain-update-perfect}
\end{equation}

\section{Rheological fracture element serial assembly}\label{sec3}

The variational formulation of the Griffith principle of fracture introduces a phase-field regularization of sharp cracks ($\Gamma$) through a crack surface density function $\gamma(l_0,\phi)$, where the scalar phase field $0 \le \phi \le 1$ represents a diffused crack over a characteristic length $l_0$. A general expression for the crack density is given by \cite{francfort_variational_2008}
\begin{equation}
  \Gamma(\phi)
  =
  \int_{\Gamma_0}
  G_c\,\mathrm{d}A_0
  \approx
  \int_{\Omega_0}
  G_c\,
  \gamma\left(l_0,\phi,\nabla_X\phi\right)
  \mathrm{d}V
  \label{eq:regularized-crack-energy}
\end{equation}
with $G_c$ [$J/m^2$] denoting the critical energy release rate.

The regularized energy functional $W(\bm{u},\phi)$ is constructed from recoverable elastic energy stored $\psi(\bm\varepsilon^e)$ and phase-field fracture energy. To suppress fracture under compressive volumetric loading, only the non-compressive part of the elastic strain energy $\psi^+$ is degraded, giving
\begin{equation}
  \bar{W}(\bm{u},\phi)
  =
  \int_{\Omega_0}
  \big[g(\phi)\psi^+ + \psi^- \big]\,\mathrm{d}V
  +
  \int_{\Omega_0} \frac{G_c}{c_0 l_0}
  \big[\alpha(\phi) + l_0^2|\nabla \phi|^2\big]\,\mathrm{d}V
  \label{eq:damage-total-energy-anisotropy}
\end{equation}
where $g(\phi)$ is the so-called degradation function and $\psi^- =\psi -\psi^+$.
Standard expressions for $\alpha(\phi)$ in Eqs. \eqref{eq:regularized-crack-energy} and \eqref{eq:damage-total-energy-anisotropy} have resulted in the Ambrosio and Tortorelli (AT1 and AT2) models \cite{tanne_crack_2018}, for which
\begin{equation}
  \alpha(\phi)=
  \begin{cases}
    \phi, & \mathrm{AT1}
    \\[6pt]
    \phi^2,
    & \mathrm{AT2}
  \end{cases}
  \label{eq:AT1-AT2}
  \end{equation}
with scaling constants $c_0 = 8/3$ and $c_0 = 2$, respectively. In the AT2 model, damage starts to accumulate as soon as the load is applied.
By contrast, the AT1 model introduces an energy threshold $\psi^{+}_c =\frac{\alpha'(\phi)G_c}{c_0l_0[-g'(0)]}$ for damage. Given the choice for the expression of the degradation function below, substituting the corresponding value of $\alpha(\phi)$ in Eq. \eqref{eq:AT1-AT2} yields $\psi^{+}_c = \frac{3G_c}{16\,l_0}$.

For the parallel rheological assembly introduced in Section 2, the stored elastic energy is obtained by summing the contributions of the individual Hooke, Maxwell, and Prandtl springs,
\begin{equation}
  \psi^\pm
  =
  \sum_k \psi_k^\pm
  \label{eq:sum_energies}
\end{equation}
where $k$ denotes an individual rheological branch. Here, we adopt the volumetric-deviatoric split of Amor \textit{et al.} \cite{amor2009regularized},
\begin{equation}
  \psi_k^+
  =
  \frac{\kappa_k}{2}
  \left\langle
    \mathrm{tr}
    (\bm\varepsilon_k^e)
  \right\rangle_+^2
  +
  \mu_k
  \bm\varepsilon_{\mathrm{d},k}^e:
  \bm\varepsilon_{\mathrm{d},k}^e
  \label{eq:positive_branch_energy}
\end{equation}
and
\begin{equation}
  \psi_k^-
  =
  \frac{\kappa_k}{2}
  \left\langle
    \mathrm{tr}
    (\bm\varepsilon_k^e)
  \right\rangle_-^2
  \label{eq:negative_branch_energy}
\end{equation}
where
$\bm\varepsilon_{\mathrm{d},k}^e
=\mathrm{dev}(\bm\varepsilon_k^e)$
and
$\langle x\rangle_\pm=(x\pm|x|)/2$.
This energy split has been shown in the literature to agree well with experimental observations in brittle fracture problems. In the present work, it is further motivated by the restriction of inelasticity to the deviatoric components, which causes damage evolution in ductile or highly viscoelastic materials to be driven primarily by the volumetric energy contributions. This is consistent with experimental observations showing that high hydrostatic pressures suppress void nucleation and coalescence in metals \cite{rice_ductile_1969}.

The degraded Kirchhoff stress follows from differentiation of the degraded elastic strain energy density with respect to the corresponding elastic Hencky strain, at fixed phase field and other branch state variables,
\begin{equation}
  \begin{aligned}
  \bm\tau_{\mathrm{degr},k}=
  \frac{\partial\psi_\mathrm{degr}}
       {\partial\bm\varepsilon_k^e}
  =
  g(\phi)
  \frac{\partial\psi^+}
       {\partial\bm\varepsilon_k^e}
  +
  \frac{\partial\psi^-}
       {\partial\bm\varepsilon_k^e}
  =
  g(\phi)
  \frac{\partial\psi_k^+}
       {\partial\bm\varepsilon_k^e}
  +
  \frac{\partial\psi_k^-}
       {\partial\bm\varepsilon_k^e}
  \end{aligned}
  \label{eq:branch_stress_from_energy}
\end{equation}
The derivatives of the positive and negative
contributions are
\begin{equation}
  \frac{\partial\psi_k^+}
       {\partial\bm\varepsilon_k^e}
  =
  \kappa_k
  \left\langle
    \mathrm{tr}
    (\bm\varepsilon_k^e)
  \right\rangle_+
  \bm I
  +
  2\mu_k\bm\varepsilon_{\mathrm{d},k}^e,
  \label{eq:positive_energy_derivative}
\end{equation}
and
\begin{equation}
  \frac{\partial\psi_k^-}
       {\partial\bm\varepsilon_k^e}
  =
  \kappa_k
  \left\langle
    \mathrm{tr}
    (\bm\varepsilon_k^e)
  \right\rangle_-
  \bm I
  \label{eq:negative_energy_derivative}
\end{equation}
Substituting into
Eq.~\eqref{eq:branch_stress_from_energy} gives
\begin{equation}
  \begin{aligned}
  \bm\tau_{\mathrm{degr},k}
  ={}&
  2g(\phi)\mu_k
  \bm\varepsilon_{\mathrm{d},k}^e
  +
  \kappa_k
  \left[
    g(\phi)
    \left\langle
      \mathrm{tr}
      (\bm\varepsilon_k^e)
    \right\rangle_+
    +
    \left\langle
      \mathrm{tr}
      (\bm\varepsilon_k^e)
    \right\rangle_-
  \right]\bm I
  \end{aligned}
  \label{eq:degraded_branch_stress}
\end{equation}
After straightforward manipulation, the updated elastic strain for the Maxwell branch in Eq. \eqref{eq:viscoelasticity-deviatoric-strain-update-hencky} becomes
\begin{equation}
\bm{\varepsilon}_{\mathrm{d},\mathrm M}^e = \left( 1 + \frac{g(\phi)\mu_{\mathrm M}\Delta t}{\eta_\mathrm{d}} \right)^{-1} 
\bm{\varepsilon}_{\mathrm{d},\mathrm M}^{e \, \mathrm{tr}}
\label{eq:viscoelasticity-deviatoric-strain-update-hencky-damage}
\end{equation}

For the Prandtl branch, the equivalent stress becomes $q^{\mathrm{tr}}(\phi)=g(\phi)q^{\mathrm{tr}}$ and the corrected deviatoric strain in a perfect plastic regime is
\begin{equation}
  \bm{\varepsilon}_{\mathrm{d,P}}^{e}
  =\left(\frac{\sigma_0}{g(\phi)q^{\mathrm{tr}}}\right)
    \bm{\varepsilon}_{\mathrm{d,P}}^{e\,\mathrm{tr}}
    \qquad \mathrm{Perfect\, plastic\, regime}.
  \label{eq:principal-directions-format-1}
\end{equation}
Notably, in the same regime,
\begin{equation}
  \bm{\tau}_{\mathrm{d, P\,degr}}
  =2 g(\phi)\mu_\mathrm{P}
  \left(\frac{\sigma_0}{g(\phi)q^{\mathrm{tr}}}\right)
  \bm{\varepsilon}_{\mathrm{d,P}}^{e\,\mathrm{tr}}
  =
  \left(\frac{\sigma_0}{q^{\mathrm{tr}}}\right)
  \bm{\tau}_{{\mathrm{d,P}}}^{\mathrm{tr}}
  \qquad \mathrm{Perfect\, plastic\, regime},
  \label{eq:degraded-perfect-plastic-stress}
\end{equation}
and the stress update is independent of damage, consistent with the proposed rheological framework in Fig. \ref{fig_rheological_model}.

Finally, the total degraded Kirchhoff stress is obtained as
\begin{equation}
  \bm\tau_{\mathrm{degr}}
  =
   \sum_k \bm\tau_{\mathrm{degr}, k}
  \label{eq:total_degraded_stress}
\end{equation}


\subsection{Residual Stiffness Factor and Damage Viscosity}
\label{sec:viscous-damage}

Quadratic, cubic, and quartic expressions for the degradation function $g(\phi)$ have been proposed in the literature \cite{kuhn2015degradation, svolos_convexity_2023}. The selected function should decrease monotonically between $g(0)=1$ and $g(1)=0$, and satisfy $g'(1)=0$. Here, we choose the standard quadratic expression

\begin{equation}
  g(\phi) = (1 - \phi)^2 + \eta
  \label{eq:degr-funct}
\end{equation}
where $\eta \ll 1$ prevents a singular Jacobian that would otherwise result from complete stiffness loss at $\phi=1$.

An additional numerical challenge arises from the nonconvexity of the coupled energy functional, which can lead to loss of stability and convergence. While arc length continuation methods have been successfully used to address this issue \cite{singh_fracture-controlled_2016, bharali_robust_2022}, they introduce additional algorithmic complexity and computational cost. Here, we adopt a viscous regularization strategy because it is well suited for inelastic materials \cite{langenfeld_how_2022} and can be implemented at minimal additional cost. Specifically, a rate-dependent term is added to the damage evolution law
\begin{equation}
  \zeta\dot{\phi} + \frac{\delta \bar W}{\delta \phi} = 0
\end{equation}
where $\zeta$ [Pa $\cdot$ s] is the damage viscosity parameter. As shown in Appendix C, this regularization adds a positive definite term to the phase-field diagonal block of the Jacobian.

The value of $\zeta$ may be chosen small enough to improve numerical convergence while preserving a rate-independent fracture response. Nevertheless, as shown below, damage viscosity promotes a more uniform distribution of damage nucleation sites and crack branching, which may lead to better agreement with experimental observations.


\subsection{Weak Form and Linearization}

In the absence of body forces, the mechanical boundary conditions are
\begin{equation}
  \bm{u}
  =
  \bar{\bm{u}}
  \quad \text{on } \Gamma_D^0,
  \qquad
  \bm{P}_{\mathrm{degr}}\bm{N}
  =
  \bar{\bm{T}}
  \quad \text{on } \Gamma_N^0
  \label{eq:mechanical-boundary-conditions}
\end{equation}
where $\bm{N}$ is the outward unit normal to the reference boundary, $\bar{\bm{T}}$ is the prescribed nominal traction,  and $\bm{P}_{\mathrm{degr}}$ is the degraded first Piola-Kirchhoff stress tensor. The natural boundary condition for the phase field is
\begin{equation}
  \nabla_X\phi\cdot\bm{N}
  =
  0
  \quad \text{on } \partial\Omega_0
  \label{eq:phase-field-boundary-condition}
\end{equation}
The displacement trial and test spaces and the phase-field space are then defined as
\begin{equation}
\begin{aligned}
  \mathcal{U}_{\bar{u}}
  &:=
  \left\{
    \bm{u}\in[H^1(\Omega_0)]^3
    \,\middle|\,
    \bm{u}=\bar{\bm{u}}
    \text{ on }\Gamma_D^0
  \right\},
  \\[6pt]
  \mathcal{V}_0
  &:=
  \left\{
    \bm{v}\in[H^1(\Omega_0)]^3
    \,\middle|\,
    \bm{v}=\bm{0}
    \text{ on }\Gamma_D^0
  \right\},
  \\[6pt]
  \mathcal{Q}
  &:=
  H^1(\Omega_0)
\end{aligned}
\label{eq:trial-test-spaces}
\end{equation}
where, $\bm{v}\in\mathcal{V}_0$ and $w\in\mathcal{Q}$ denote the displacement and phase-field test functions, respectively. The variational problem consists of finding
\begin{equation}
  (\bm{u},\phi)
  \in
  \mathcal{U}_{\bar{u}}\times\mathcal{Q}
\end{equation}
such that
\begin{equation}
\begin{aligned}
  r_u
  &:=
  \int_{\Omega_0}
    \nabla_x\bm{v}:
    \bm{\tau}_{\mathrm{degr}}
  \,\mathrm{d}V
  -
  \int_{\Gamma_N^0}
    \bm{v}\cdot\bar{\bm{T}}
  \,\mathrm{d}A
  =
  0,
  &&
  \forall\bm{v}\in\mathcal{V}_0,
  \\[6pt]
  r_\phi
  &:=
  \int_{\Omega_0}
  \left[
    wL
    +
    \nabla_xw\cdot
    \frac{2G_cl_0}{c_0}
    \bm{b}\nabla_x\phi
  \right]
  \mathrm{d}V
  =
  0,
  &&
  \forall w\in\mathcal{Q}
\end{aligned}
\label{eq:hyperelastic-damage-weak-form-current}
\end{equation}

In Eq. \eqref{eq:hyperelastic-damage-weak-form-current}, the weak forms are integrated over the reference configuration but expressed in terms of spatial gradients and the Kirchhoff stress, using the identities
\begin{equation}
  \nabla_X\bm{v}:\bm{P}_{\mathrm{degr}}
  =
  \nabla_x\bm{v}:\bm{\tau}_{\mathrm{degr}}.
  \label{eq:mechanical-gradient-transformation}
\end{equation}
and
\begin{equation}
\begin{aligned}
  \nabla_X w\cdot\nabla_X\phi
  =
  \left(\bm{F}^{T}\nabla_x w\right)
  \cdot
  \left(\bm{F}^{T}\nabla_x\phi\right)
  =
  \nabla_x w\cdot
  \bm{F}\bm{F}^{T}\nabla_x\phi
  =
  \nabla_x w\cdot\bm{b}\nabla_x\phi
\end{aligned}
\label{eq:phase-gradient-transformation}
\end{equation}
The contribution $L$ to the phase field residual is instead
\begin{equation}
  L(\bm{u},\phi,\dot{\phi})
  =
  g'(\phi)\mathcal{H}
  +
  \frac{G_c}{c_0l_0}\alpha'(\phi)
  +
  \zeta\dot{\phi}
  \label{eq:L-viscous}
\end{equation}
where $\mathcal{H}$ is the history variable introduced in Miehe \textit{et al.} \cite{miehe_phase_2010} to enforce crack irreversibility by setting $\mathcal{H}_{n+1}=\max\left(\mathcal{H}_n, \psi_{n+1}^{+}\right)$. 

Let $(\delta\bm{u},\delta\phi)\in\mathcal{V}_0\times\mathcal{Q}$ denote an admissible solution increment, the consistent directional linearization of Eq. \eqref{eq:hyperelastic-damage-weak-form-current} gives
\begin{equation}
\begin{aligned}
  \mathrm{D}r_u
  (\bm{u},\phi,\bm{v})
  [\delta\bm{u},\delta\phi]
  =
  \int_{\Omega_0}
  \nabla_x\bm{v}:
  \left[
    \delta\bm{\tau}_{\mathrm{degr}}
    -
    \bm{\tau}_{\mathrm{degr}}
    \left(
      \nabla_x\delta\bm{u}
    \right)^{T}
  \right]
  \mathrm{d}V
  =
  -r_u,
  &&
  \forall\bm{v}\in\mathcal{V}_0,
  \\[6pt]
  \mathrm{D}r_\phi
  (\bm{u},\phi,w)
  [\delta\bm{u},\delta\phi]
  =
  \int_{\Omega_0}
  \left[
    w\,\delta L
    +
    \nabla_xw\cdot
    \frac{2G_cl_0}{c_0}
    \bm{b}\nabla_x\delta\phi
  \right]
  \mathrm{d}V
  =
  -r_\phi,
  &&
  \forall w\in\mathcal{Q}
\end{aligned}
\label{eq:hyperelastic-damage-linearization-current}
\end{equation}
where $\delta\bm{\tau}_{\mathrm{degr}}$ and $\delta L$ respectively denote the directional linearizations of the degraded Kirchhoff stress and the local phase-field residual, as derived in Appendix C.

\section{Matrix-free implementation}
\label{sec4}

\begin{figure*}[!h]%
  \centerline{\includegraphics[width=450pt,height=13pc, trim=5 5 8 5,clip]{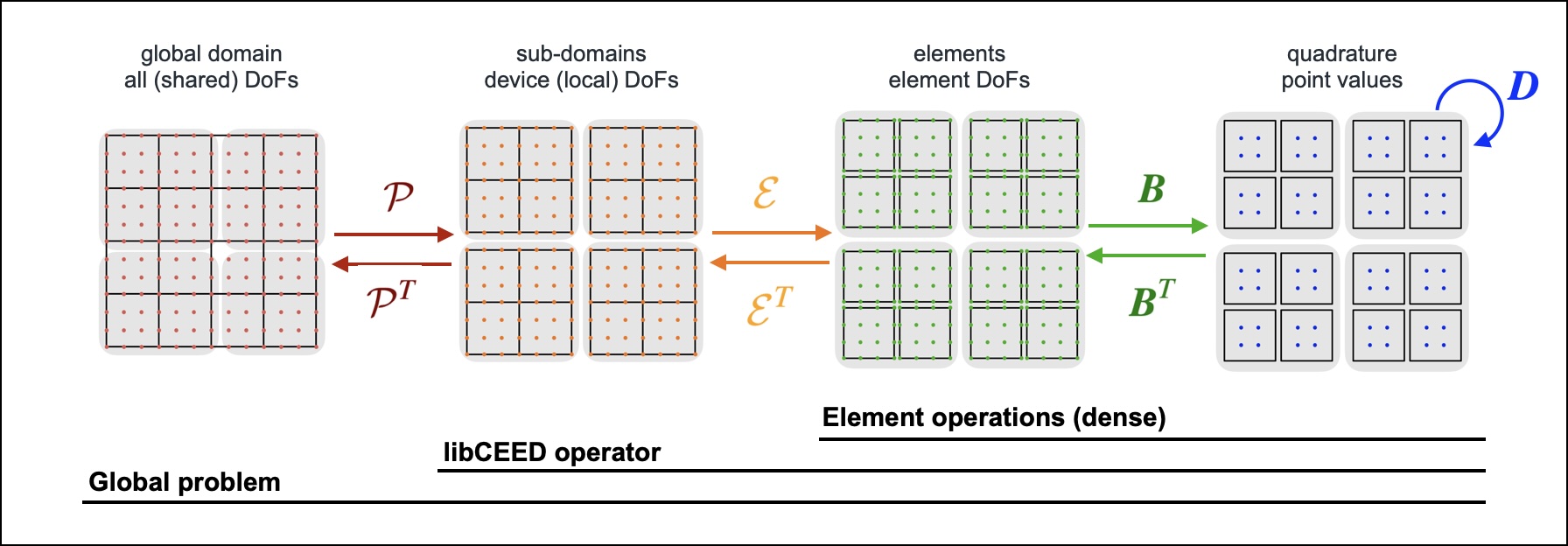}}
  \caption{Matrix-free operator decomposition in libCEED. Adapted from Brown \textit{et al.} \cite{brown_performance_2022}. Note that the sub-domains vector (L-vector in Brown \textit{et al.}) is localized per device with ghost values replicated into each part. \label{fig_libceed_operator}}
  \end{figure*}

The coupled phase-field formulation is implemented using the matrix-free finite element operator abstraction provided by libCEED and described in Brown \textit{et al.} \cite{brown_performance_2022}. As illustrated in Fig. \ref{fig_libceed_operator}, the action of a finite-element operator can be expressed as
\begin{equation}
  \boldsymbol{A}
  =
  \color{Maroon}{\mathcal{P}^{T}}
  \color{Orange}{\mathcal{E}^{T}}
  \color{Green}{\boldsymbol{B}^{T}}
  \color{Blue}{\boldsymbol{D}}
  \color{Green}{\boldsymbol{B}}
  \color{Orange}{\mathcal{E}}
  \color{Maroon}{\mathcal{P}}\color{black}
  \label{eq:libceed-operator}
\end{equation}
where $\color{Maroon}\mathcal{P}$ handles parallel communication between global and process-local degrees of freedom, $\color{Orange}\mathcal{E}$ restricts local degrees of freedom to elements, and $\color{Green}\boldsymbol{B}$ evaluates the finite element basis functions and their derivatives at quadrature points. The operator $\color{Blue}\boldsymbol{D}$ represents the pointwise evaluation of the weak form integrands and their linearizations, including constitutive responses and geometric transformations.
This decomposition separates the finite element basis operations from the material specific computations, allowing the same matrix-free operator structure to accommodate the different rheological branches and the phase-field fracture model. The global Jacobian matrix is not explicitly assembled. Instead, its action is computed through the corresponding element-local operations.

Let $\bm{z}=[u_x,u_y,u_z,\phi]^T$ denote the solution vector and let $\bm{\xi}$ denote the coordinates of the parent finite element. The volume contribution to the weak form residual in Eq. \eqref{eq:hyperelastic-damage-weak-form-current} can be written as
\begin{equation}
  r_{\Omega}(\bm{z};\bm{v},w)
  =
  \int_{\Omega_0}
  \left[
    \bm{q}\cdot\bm{f}_0
    +
    \nabla_x\bm{q}:\bm{f}_1
  \right]
  \mathrm{d}V
  \label{eq:matrix-free-weak-form}
\end{equation}
where $\bm{q}=[v_x,v_y,v_z,w]^T$ collects the displacement and phase-field test functions. The pointwise integrands are
\begin{equation}
  \bm{f}_0
  =
  \begin{bmatrix}
    \bm{0}\\
    s_{\phi}L
  \end{bmatrix},
  \qquad
  \bm{f}_1
  =
  \begin{bmatrix}
    \bm{\tau}_{\mathrm{degr}}\\[6pt]
    s_{\phi}
    \dfrac{2G_cl_0}{c_0}
    \bm{b}\nabla_x\phi
  \end{bmatrix}
  \label{eq:physical-residual-integrands}
\end{equation}
where $s_{\phi}>0$ is a scaling factor applied to the phase-field residual to improve the relative scaling of the coupled equations.
The prescribed mechanical traction contributes separately through the boundary integral in Eq.~\eqref{eq:hyperelastic-damage-weak-form-current}.

The spatial gradients appearing in the quadrature point integrands are expressed in terms of derivatives with respect to the parent element coordinates. By the chain rule,
\begin{equation}
  \nabla_x\bm{z}
  =
  \nabla_{\xi}\bm{z}\,
  \nabla_x\bm{\xi}
  \label{eq:parent-spatial-gradient}
\end{equation}
The transformation is related to the reference and current configurations by
\begin{equation}
  \nabla_x\bm{\xi}
  =
  \nabla_X\bm{\xi}\bm{F}^{-1}
  \label{eq:parent-current-mapping}
\end{equation}
Thus, the reference element basis derivatives can be evaluated independently of the solution-dependent transformation to spatial
coordinates.

For element $el$, the values and parent coordinate gradients of the solution at quadrature points are obtained as
\begin{equation}
  \bm{z}_{el}
  =
  \bm{B}_I
  \mathcal{E}_{el}
  \mathcal{P}\bm{z},
  \qquad
  \bm{G}_{\xi,el}
  =
  \bm{B}_{\xi}
  \mathcal{E}_{el}
  \mathcal{P}\bm{z}
  \label{eq:element-quadrature-fields}
\end{equation}
where the operators $\bm{B}_I$ and $\bm{B}_{\xi}$ evaluate the finite element basis functions and their parent coordinate derivatives at quadrature points, respectively.

The physical integrands in Eq. \eqref{eq:physical-residual-integrands}
are transformed to parent element coordinates as
\begin{align}
  \hat{\bm{f}}_0
  \left(
    \bm{z}_{el},
    \bm{G}_{\xi,el}
  \right)
  &=
  \bm{f}_0
  \left(
    \bm{z}_{el},
    \bm{G}_{\xi,el}\nabla_x\bm{\xi}
  \right),
  \label{eq:f0-reference}
  \\[6pt]
  \hat{\bm{f}}_1
  \left(
    \bm{z}_{el},
    \bm{G}_{\xi,el}
  \right)
  &=
  \bm{f}_1
  \left(
    \bm{z}_{el},
    \bm{G}_{\xi,el}\nabla_x\bm{\xi}
  \right)
  \left(
    \nabla_x\bm{\xi}
  \right)^T
  \label{eq:f1-reference}
\end{align}
The resulting discrete volume residual is
\begin{equation}
\begin{aligned}
  \bm{R}_{\Omega}(\bm{z})
  =
  \mathcal{P}^{T}
  \sum_{el}
  \mathcal{E}_{el}^{T}
  \begin{bmatrix}
    \bm{B}_I\\
    \bm{B}_{\xi}
  \end{bmatrix}^{T}
  \bm{W}_{el}\bm{\Lambda}_{el}
  \begin{bmatrix}
    \hat{\bm{f}}_0
    \left(
      \bm{z}_{el},
      \bm{G}_{\xi,el}
    \right)\\[6pt]
    \hat{\bm{f}}_1
    \left(
      \bm{z}_{el},
      \bm{G}_{\xi,el}
    \right)
  \end{bmatrix}
\end{aligned}
\label{eq:matrix-free-residual-compact}
\end{equation}
Where $\bm{W}_{el}\bm{\Lambda}_{el}$ represents the quadrature weights multiplied by the geometric factors of the mapping from the parent element to the reference configuration. The transpose operators perform the corresponding accumulation of quadrature contributions into element,
process-local, and global vectors.

The Jacobian vector product is evaluated using the same element-local operator structure:
\begin{equation}
\begin{aligned}
  \bm{J}(\bm{z})\delta\bm{z}
  =
  \mathcal{P}^{T}
  \sum_{el}
  \mathcal{E}_{el}^{T}
  \begin{bmatrix}
    \bm{B}_I\\
    \bm{B}_{\xi}
  \end{bmatrix}^{T}
  \bm{W}_{el}\bm{\Lambda}_{el}
  \begin{bmatrix}
    \hat{\bm{f}}_{0,0}
      &
    \hat{\bm{f}}_{0,1}\\
    \hat{\bm{f}}_{1,0}
      &
    \hat{\bm{f}}_{1,1}
  \end{bmatrix}
  \begin{bmatrix}
    \bm{B}_I\\
    \bm{B}_{\xi}
  \end{bmatrix}
  \mathcal{E}_{el}
  \mathcal{P}\delta\bm{z}
\end{aligned}
\label{eq:matrix-free-jacobian}
\end{equation}
The pointwise Jacobian blocks are defined by
\begin{equation}
  \hat{\bm{f}}_{i,0}
  :=
  \frac{\partial\hat{\bm{f}}_i}
       {\partial\bm{z}_{el}},
  \qquad
  \hat{\bm{f}}_{i,1}
  :=
  \frac{\partial\hat{\bm{f}}_i}
       {\partial\bm{G}_{\xi,el}},
  \qquad
  i\in\{0,1\}
  \label{eq:matrix-free-pointwise-jacobian}
\end{equation}
These derivatives include the constitutive and geometric contributions associated with the consistent linearization of the coupled weak form described in previous sections.


\section{Numerical Experiments}\label{sec5}

\begin{figure*}[!h]%
  \centerline{\includegraphics[width=450pt,height=15pc, trim=5 5 8 5,clip]{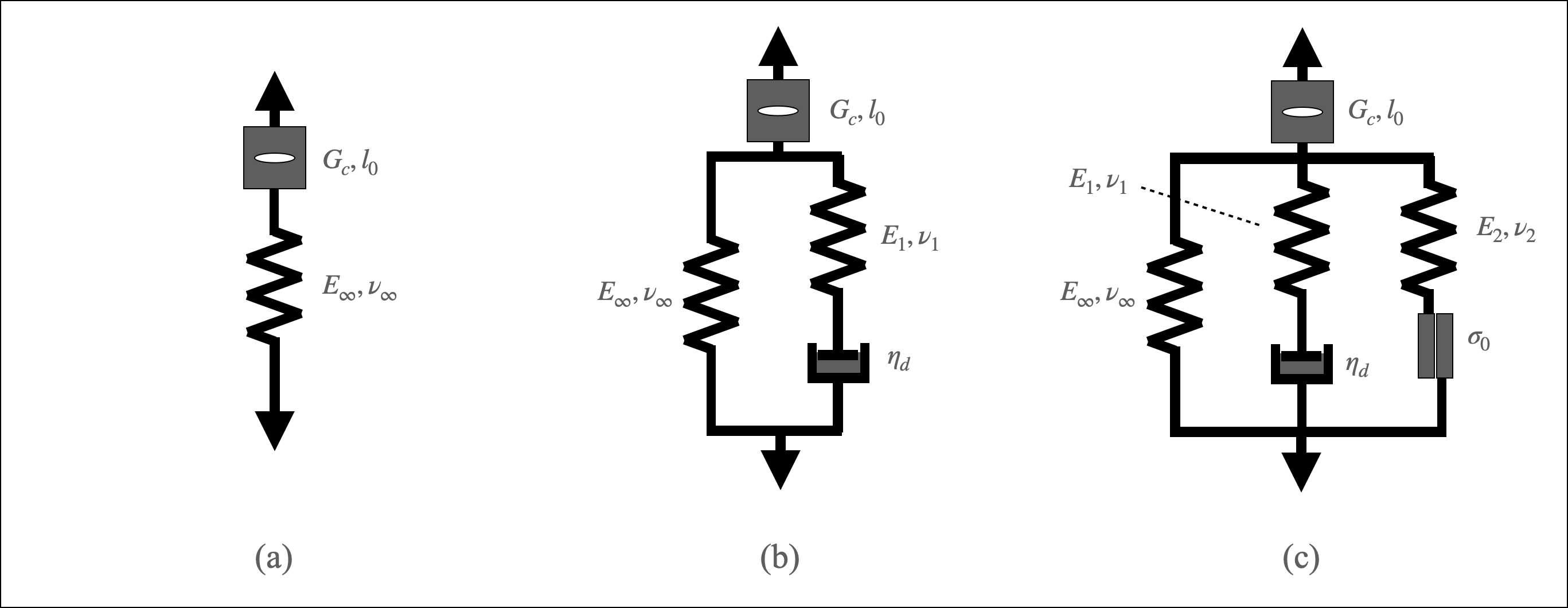}}
  \caption{Rheological elements configurations considered in the present study for: (a) notched steel plate (Example 1) and particles phase in Examples 2 and 3, (b) viscoelastic matrix (Example 2), and (c) visco-elastoplastic matrix (Example 3).\label{fig_rheological_cases}}
  \end{figure*}

We consider three numerical experiments (Examples 1-3) with rheological assemblies for the different phases illustrated in Fig. \ref{fig_rheological_cases} and values of material parameters for Examples 2 and 3 in Table \ref{table_mat_par}:

\begin{enumerate}

\item Shearing of a sharp-notched steel plate in which a single crack is expected to initiate and propagate from the notch. The rheological model is the brittle fracture model in Fig. \ref{fig_rheological_cases}a, with the rheological fracture element described using AT2. The values of the material parameters were taken from the literature: $E_\infty = 210 \, \mathrm{GPa}$ , $\nu_\infty = 0.3$, $G_c =2.7 \, \mathrm{kJ/m^2}$, and $l_0 =0.01 \, \mathrm{mm}$. This benchmark test highlights key features of phase-field fracture models, including crack insensitivity to compressive stress states. It is also commonly used to assess a monolithic solver's ability to overcome the loss of convexity in the coupled system. In the final stage of crack propagation leading to full fracture, predictions of the crack path and load-displacement curve differ in the literature (e.g., compare results in Gerasimov \& Lorenzis \cite{gerasimov_line_2016} with Kristensen \& Martinez-Paneda \cite{kristensen_phase_2020}). Investigating this aspect is beyond the scope of the present study. Nevertheless, we propose modified boundary conditions that may yield a more deterministic crack response. We then use the test to estimate sufficiently high values of the residual stiffness factor and damage viscosity to prevent numerical instabilities.
\item Quasistatic uniaxial compression of a cylindrical specimen with a relatively high volume fraction of hard ellipsoidal particles in a viscoelastic matrix. This configuration is representative of micromechanical testing of granular materials, such as polymer-bonded explosives (PBX) materials \cite{manner_situ_2017}, which show a quasi-brittle response at room temperature with peak stresses within 1-2\% applied strain. We model the particle phase using the brittle fracture rheological element shown in Fig. \ref{fig_rheological_cases}a, while the softer, less fracture-resistant matrix phase is described by the viscoelastic model shown in Fig. \ref{fig_rheological_cases}b. We use the AT2 fracture model for both phases. In particular, we calibrate the matrix fracture toughness so that complete structural failure occurs at an applied strain below 1\%.
We compare the mechanical responses under two loading conditions: constrained lateral displacement at the top and bottom surfaces and lubricated contact, with the latter using contact algorithms already implemented in \texttt{Ratel}.
\item Quasistatic uniaxial tension of a rod containing a relatively low volume fraction of hard spherical particles in a viscoplastic metallic matrix. As in the previous example, the particle phase follows the brittle fracture model shown in Fig. \ref{fig_rheological_cases}a, while the softer matrix phase follows the visco-elastoplastic rheological model shown in Fig. \ref{fig_rheological_cases}c. This configuration represents micromechanical testing of particle-reinforced alloys, with a well-documented sequence of void nucleation around stiff inclusions, void growth and coalescence, and concurrent necking leading to shear banding and cup-and-cone type crack patterns \cite{tvergaard_analysis_1984}. Because crack nucleation in these materials occurs in the later stages of applied deformation, we adopt the AT1 fracture model for both phases. In particular, we calibrate the matrix fracture toughness so that appreciable necking develops before complete structural failure.

\end{enumerate}

\begin{table*}[!h]
  \centering
  \caption{Material parameters for the synthetic particle-matrix microstructures in Examples 2 and 3.}\label{table_mat_par}
  
  \renewcommand{\arraystretch}{1.15}
  
  \begin{tabular}{l l l c c c}
  \toprule
  \textbf{Phase} & \textbf{Element type} & \textbf{Parameter} & \textbf{Units} &
  \textbf{Example 2} & \textbf{Example 3}\\
  \midrule
  
  \multirow{4}{*}{\textbf{Particle}}
  & \multirow{2}{*}{Fracture} 
  & $G_c$    & kJ\,m$^{-2}$ & $0.1$ & $8$ \\
  &         & $l_0$   & mm      & $0.01$ & $0.005$ \\
  \cmidrule(lr){2-6}
  
  & \multirow{2}{*}{Hooke} 
  & $E_\infty$ & GPa & $10$ & $400$ \\
  &            & $\nu_\infty$ & -- & $0.3$ & $0.25$ \\
  
  \midrule
  
  \multirow{9}{*}{\textbf{Matrix}}
  & \multirow{2}{*}{Fracture} 
  & $G_c$    & kJ\,m$^{-2}$ & $0.01$ & $0.8$ \\
  &         & $l_0$   & mm      & $0.01$ & $0.005$ \\
  \cmidrule(lr){2-6}
  
  & \multirow{2}{*}{Hooke} 
  & $E_\infty$ & GPa & $1.0$ & $1.0$ \\
  &            & $\nu_\infty$ & -- & $0.4$ & $0.3$ \\
  \cmidrule(lr){2-6}
  
  & \multirow{3}{*}{Maxwell} 
  & $E_1$ & GPa & $1.0$ & $1.0$ \\
  &       & $\nu_1$ & -- & $0.4$ & $0.3$ \\
  &       & $\eta_d$ &GPa $\cdot$ s & $0.1, +\infty$ & $0.01$ \\
  \cmidrule(lr){2-6}
  
  & \multirow{3}{*}{Prandtl -- Perfect Pl.} 
  & $E_2$ & GPa & -- & $198$ \\
  &       & $\nu_2$ & -- & -- & $0.30$ \\
  &       & $\sigma_0$ & MPa & -- & $500$ \\
  \bottomrule
  \end{tabular}
  \end{table*}
  
Synthetic particle-matrix microstructures were generated using \texttt{Python} scripts that leverage \texttt{microstructpy} \cite{hart_microstructpy_2020} to control the shape, size, and volume fraction of particles. In particular, the ratios of the minor to major axes and the particle sizes were drawn from lognormal distributions. Based on a user provided random number generator, \texttt{microstructpy} uniformly distributes particles within a unit cube and stores the positions and shapes of the individual particles in a seed list. The scripts read into the seed list to mask out particles that lie outside a user-provided cylindrical volume.

For ellipsoidal particles (Example 2), the geometries were converted to Stereolithography (or Standard Triangle Language) STL surfaces using \texttt{trimesh} \cite{dawson2019trimesh}, and potential overlaps between STLs were filtered. Meshing of both the particle and matrix phases was performed using \texttt{Gmsh} \cite{geuzaine2009gmsh} and subsequently optimized with \texttt{Netgen} \cite{schoberl1997netgen} to ensure high mesh quality. Git repositories \cite{DiGioacchino_micromorph_2024, DiGioacchino_micromorph_2025} include the documented Python scripts used here to generate, post-process, and mesh the synthetic microstructures and sample volumes.

Numerical simulations were performed using quadratic (P2) finite elements. The nonlinear systems were solved using a Newton-Raphson method with backtracking line search and adaptive time stepping. At each Newton iteration, the linearized systems were solved using GMRES with adaptive classical Gram-Schmidt refinement to maintain Krylov basis orthogonality. Preconditioning was performed using $p$-multigrid with algebraic multigrid (AMG) coarse solvers and Chebyshev smoothing. Since the characteristic length scale is similar to the mesh size, the damage field was excluded from the near-null space provided to AMG, thus decreasing the grid complexity and improving performance. In Example 3, in particular, the coarse level adopted Hypre BoomerAMG \cite{falgout_hypre_2002}, as it resulted in a reduced number of linear solver iterations. 

All simulations were run on the Tioga high-performance computing cluster at Lawrence Livermore National Laboratory, an AMD-based GPU system running the GFX90a architecture. Each Tioga node contains 4 AMD Instinct MI250X GPUs, each of which has 128 GB of HBM2e device memory and is split into two logical GPUs, and 64 AMD EPYC CPU cores with 512 GB of host memory. Via PETSc\cite{petsc_exascale_2025}, the code was compiled using Kokkos Kernels \cite{trott_kokkos_2022}, with HIP serving as the backend for execution on AMD GPUs. The \texttt{libCEED}\cite{Brown2021} code generation backend for HIP, \texttt{/gpu/hip/gen}, was utilized for matrix-free operator application and assembly. All simulations can be reproduced on CPU- or CUDA-based systems via the corresponding PETSc and \texttt{libCEED} backends. The microstructure-resolved simulations of Examples 2 and 3 resulted in problem sizes of up to approximately 20 million DoFs, and were completed in about 500 time steps using up to 3 nodes, 24 logical GPUs, and within a 12-hour allocation window.

\subsection{Sharp-notched Plate Shearing Test}

\begin{figure*}[!h]%
    \centerline{\includegraphics[width=465pt,height=13pc, trim=5 5 8 5,clip]{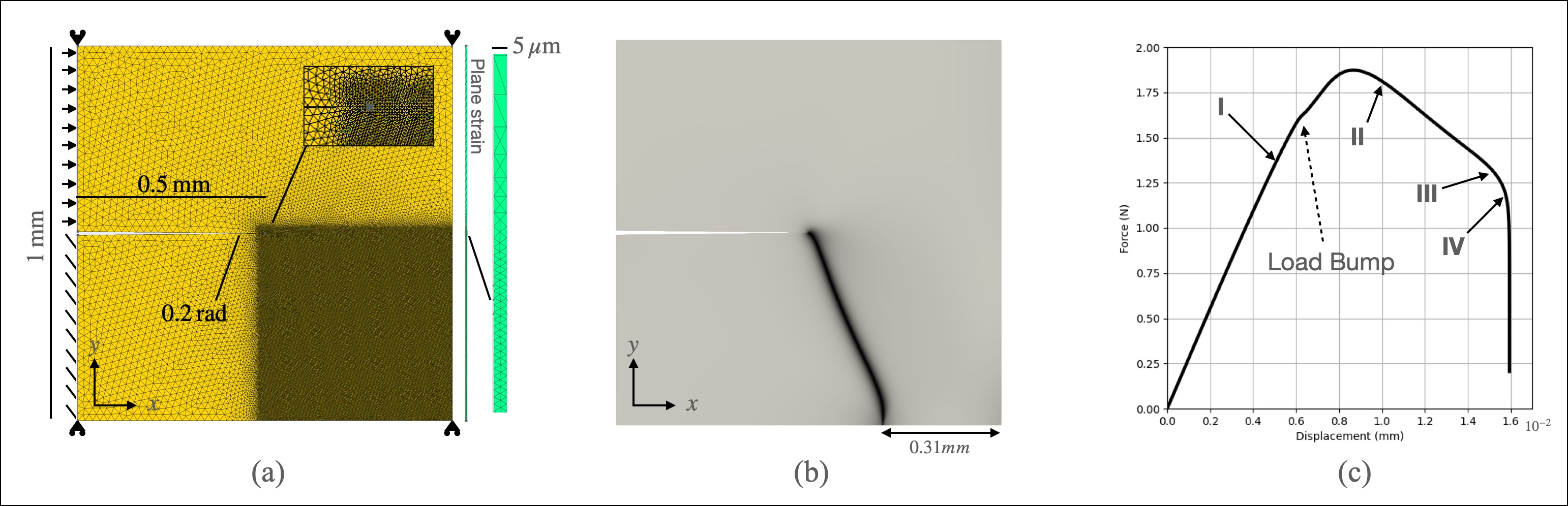}}
    \caption{(a) FE mesh and boundary conditions for the sharp-notched plate shearing test. (b) Grayscale plot to visualize the full crack path. (c) Force-displacement curve with highlighted stages for which convergence rates are reported in Table \ref{tab_convergence_rates}.\label{fig_plate_1}}
    \end{figure*}

Figure \ref{fig_plate_1}a shows the three-dimensional finite element mesh of the sharp-notched plate for Example 1, including local refinement in the region where crack propagation is expected. The plane strain boundary conditions mimic those found in the literature, and in which simple shear is applied on the plate by clamping its bottom surface and imposing a lateral displacement on its top surface. To allow crack opening at the boundary, we modify the boundary conditions so that these surfaces are simply supported, and the applied shear strain arises from the relative displacement of the lateral faces, as depicted in the same figure.

Consistent with results from 2D models and small strains formulations in the literature, the crack in Fig. \ref{fig_plate_1}b initiates at the notch and its nucleation coincides with the first peak of the characteristic bump in the load-displacement curve highlighted in Fig. \ref{fig_plate_1}c. The crack propagates within the tensile region at an angle to the edges of the plate, approaching the bottom of the plate. This intermediate stage of crack propagation is accompanied by a gradual decrease in load. In the literature, the crack path is predicted to coast the clamped bottom at an infinitesimal distance. In the simply supported face considered here, the main crack path is instead predicted to bend downward, reaching the bottom surface at a right angle and at about $0.31\, \mathrm{mm}$ from the nearest corner.

Convergence rates of the nonlinear solver for Example 1 are reported for representative stages of crack propagation in Table \ref{tab_convergence_rates}. The quadratic convergence achieved during damage accumulation (I in the table) and crack nucleation at the notch tip is observed to degrade during crack propagation (II and III). In the final stage of crack propagation (IV), the solver exhibits an initial phase of slow residual reduction before entering the local convergence basin. This is associated with progressive unstable crack propagation conditions near the final ligament to complete fracture.

\begin{figure*}[!h]%
    \centerline{\includegraphics[width=460pt,height=13pc, trim=5 5 8 5,clip]{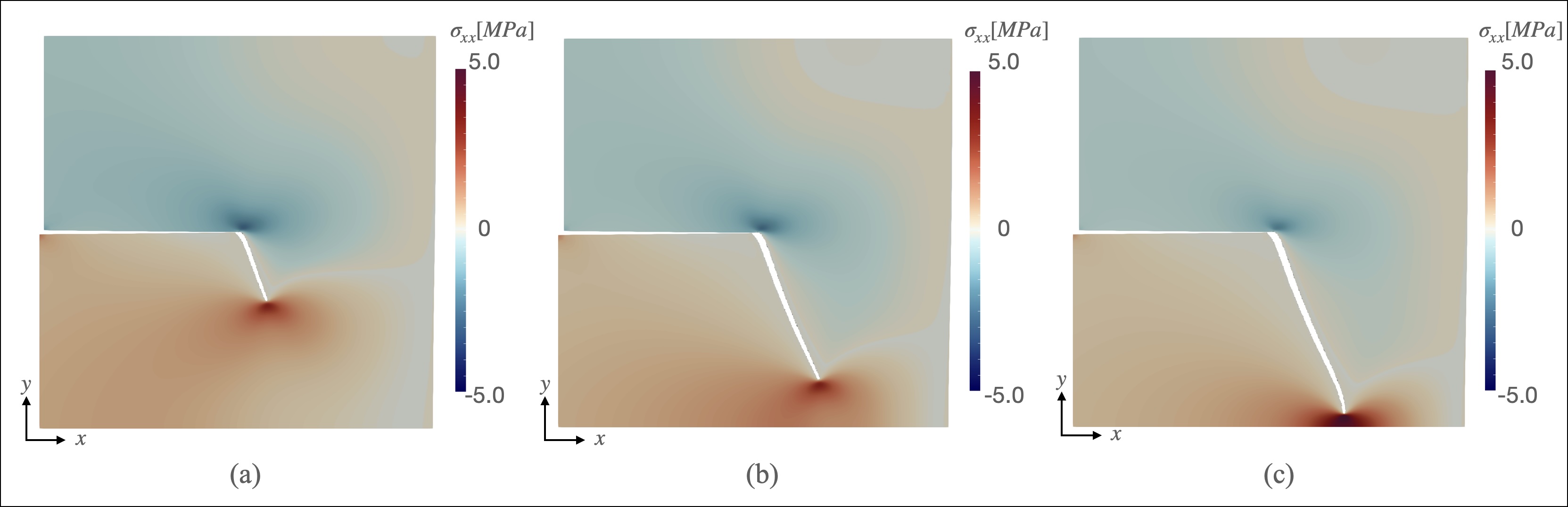}}
    \caption{Evolution of $\sigma_{xx}$ during crack propagation for three successive stages of loading. Crack opening is visualized by eliminating regions of damage field values above 0.95. \label{fig_plate_2}}
    \end{figure*}

Cutting through the final ligament would exhibit significant snapback behavior, which can be inferred from the evolution of the stress at the crack tip in Fig. \ref{fig_plate_2}. In the crack propagation stage that follows crack initiation (Figs. \ref{fig_plate_2}a and \ref{fig_plate_2}b), the stress remains at a comparable level, but approximately doubles at the final ligament, indicating a propensity for unstable crack growth. Numerical instability and potential loss of solver convergence are prevented by prescribing residual stiffness factor and damage viscosity of $\eta = 1 \cdot 10^{-3}$ and $ \zeta= 1 \cdot 10^{-2}$ MPa $\cdot$ s, respectively. These values were used in Example 3. In Example 2, the residual stiffness factor was increased to $\eta = 1 \cdot 10^{-2}$ to maintain numerical stability in the lubricated platens case, which introduces additional nonlinearity due to contact.

\begin{table*}[!h]
  \centering
  \caption{Residual function norm convergence history at the four representative stages of crack propagation highlighted in Fig. \ref{fig_plate_1}c and obtained with same time step size.}
  
  \label{tab_convergence_rates}
  \renewcommand{\arraystretch}{1.15}
  \begin{tabular}{l c c c c}
  \toprule
  \textbf{Newton iter.} $k$ &
  \textbf{I}($u_x=0.005$) &
  \textbf{II}($u_x=0.010$) &
  \textbf{III}($u_x=0.015$) &
  \textbf{IV}($u_x=0.0157$) \\
  \midrule
  0  & $7.047\times10^{-1}$ & $7.038\times10^{-1}$ & $7.029\times10^{-1}$ & $7.027\times10^{-1}$ \\
  1  & $4.257\times10^{-3}$ & $5.214\times10^{-3}$ & $5.496\times10^{-3}$ & $7.386\times10^{-3}$ \\
  2  & $4.161\times10^{-6}$ & $2.774\times10^{-3}$ & $3.028\times10^{-3}$ & $5.679\times10^{-3}$ \\
  3  & $4.021\times10^{-11}$& $8.390\times10^{-5}$ & $1.202\times10^{-4}$ & $3.116\times10^{-3}$ \\
  4  & --                   & $2.877\times10^{-6}$ & $5.540\times10^{-7}$ & $1.903\times10^{-3}$ \\
  5  & --                   & $3.849\times10^{-10}$& $2.924\times10^{-11}$& $5.949\times10^{-5}$ \\
  6  & --                   & --                   & --                   & $3.473\times10^{-7}$ \\
  7  & --                   & --                   & --                   & $5.008\times10^{-12}$ \\
  \bottomrule
  \end{tabular}
\end{table*}

\subsection{Viscoelastic Granular Material in Compression}

\begin{figure*}[!h]%
  \centerline{\includegraphics[width=450pt,height=15pc, trim=5 5 8 5,clip]{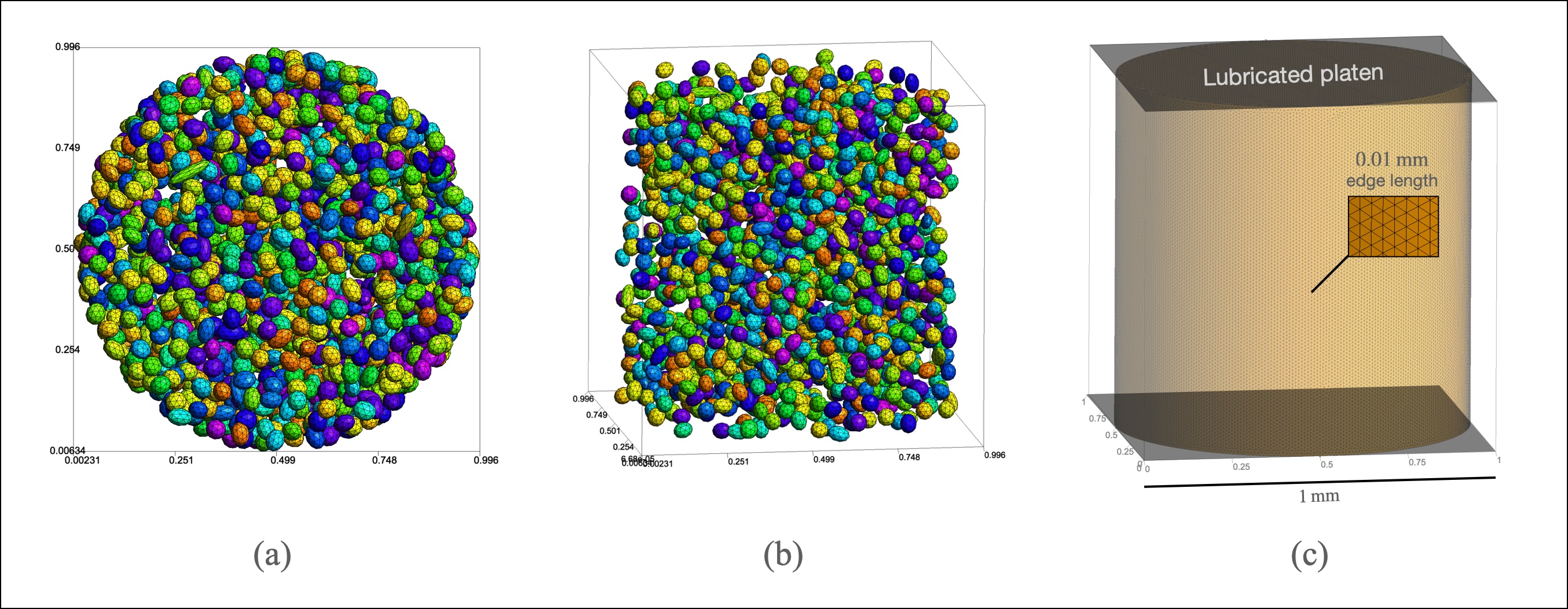}}
  \caption{Granular viscoelastic material deformed in compression. 3D rendering of (a) particles (top view), (b) particles (side view), and (c) full specimen mesh.\label{fig_granular_1}}
  \end{figure*}

Figures \ref{fig_granular_1}a and \ref{fig_granular_1}b show the 1498 ellipsoidal grains generated for Example 2, with a mean grain size of $0.02\, \mathrm{mm}$ to collectively occupy about 0.2 of the cylindrical volume. The corresponding finite element mesh, generated with a maximum element size of $0.01 \, \mathrm{mm}$, is shown in Fig. \ref{fig_granular_1}c.

\begin{figure*}[!h]%
  \centerline{\includegraphics[width=450pt,height=15pc, trim=5 5 8 5,clip]{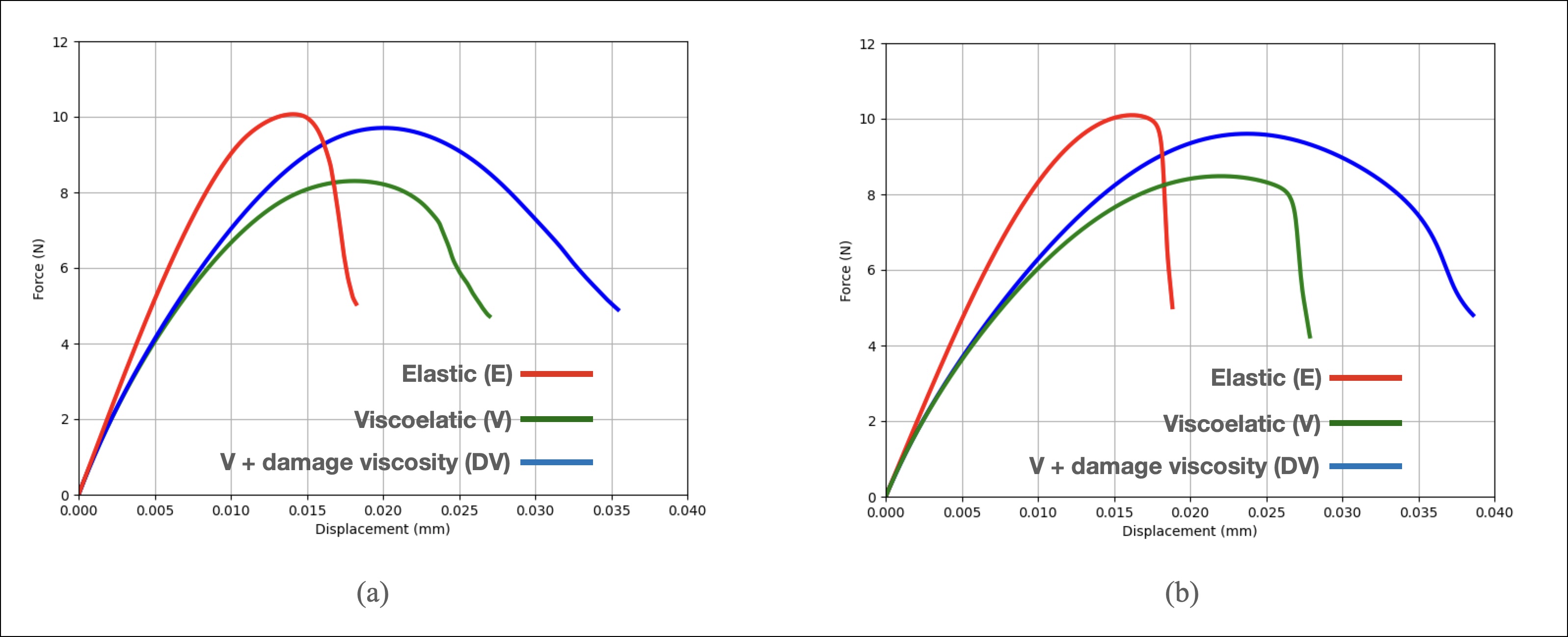}}
  \caption{Granular viscoelastic material deformed in compression. Force-displacement responses for three matrix material models evaluated under lubricated and laterally constrained conditions. (a) Laterally constrained case. (b) Lubricated platens.\label{fig_granular_2}}
  \end{figure*}

The cylindrical volumes are loaded in unconfined compression along their axis at a strain rate of $5.0 \times 10^{-2} \, \mathrm{s}^{-1}$. Lubrication is introduced through viscous damping, such that higher transverse velocities lead to increased resistance to lateral displacement at the contact surfaces (see \texttt{Ratel} documentation \cite{atkins_ratel_2026} for details). The value of the viscous damping coefficient is chosen sufficiently large to suppress convergence issues associated with near-null space modes. Three matrix materials are considered for both lubricated and laterally constrained cases: purely elastic, which is identified below as (E), finite deviatoric viscosity (VE), and finite deviatoric viscosity combined with increased damage viscosity (DV). The predicted force-displacement curves for the laterally constrained and lubricated cases are shown in Figs. \ref{fig_granular_2}a and \ref{fig_granular_2}b, respectively. These are plotted to an applied post peak displacement that corresponds to half the peak force, beyond which the relatively high residual stiffness factor adopted in this example tends to inhibit further crack opening and the associated additional force drop. 

\begin{figure*}[!h]%
    \centerline{\includegraphics[width=500pt,height=20pc, trim=5 5 8 8,clip]{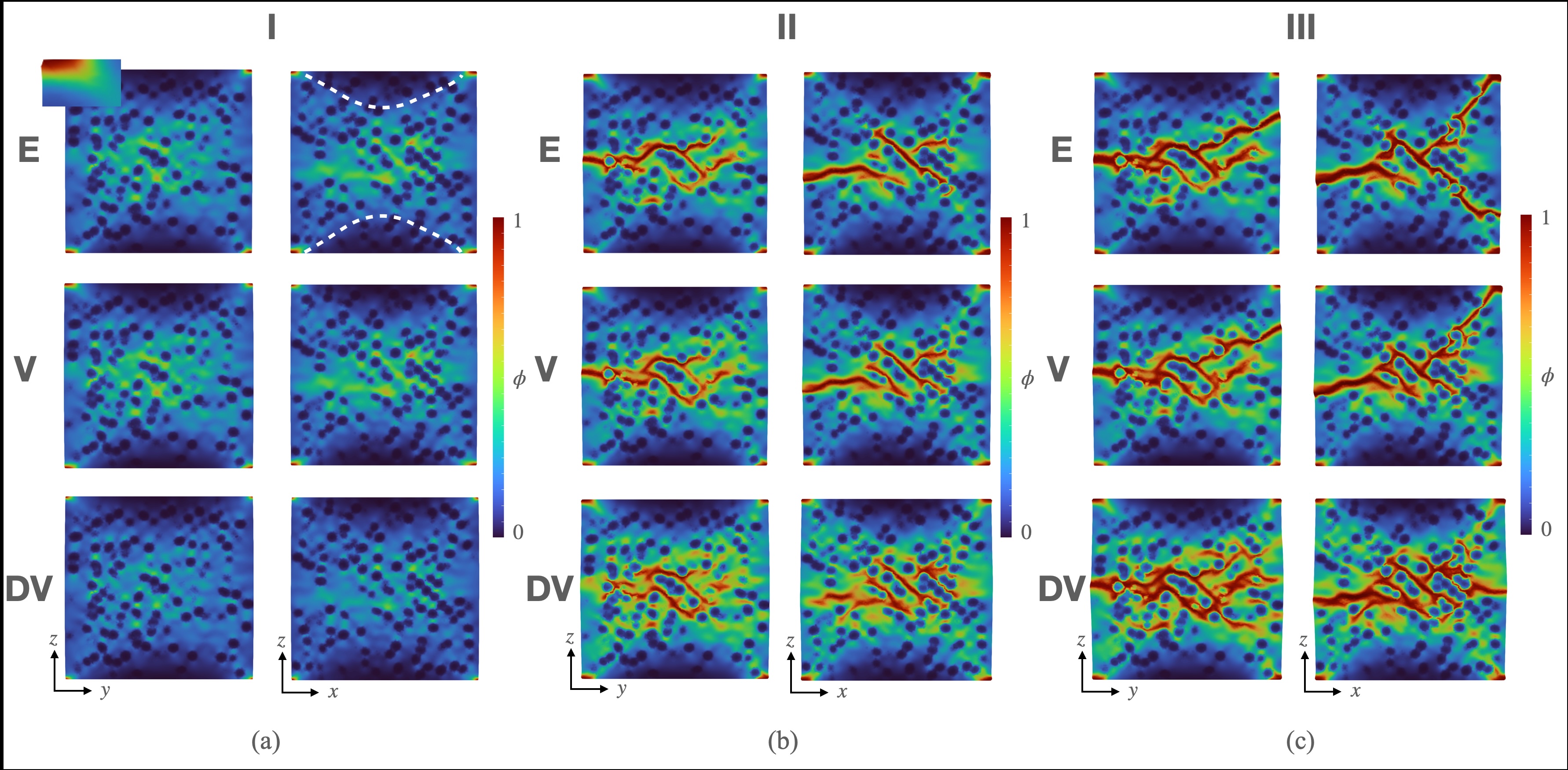}}
    \caption{Granular viscoelastic material deformed in compression, laterally constrained case. Evolution of the phase-field damage variable on $yz$ and $xz$ central cross-sections of the specimen at increasing deformation levels. (a) Peak load (I), (b) post peak to 75\% peak load (II), and (c) 50\% peak load (III).\label{fig_granular_3}}
    \end{figure*}

\begin{figure*}[!h]%
      \centerline{\includegraphics[width=500pt,height=20pc, trim=5 5 8 8,clip]{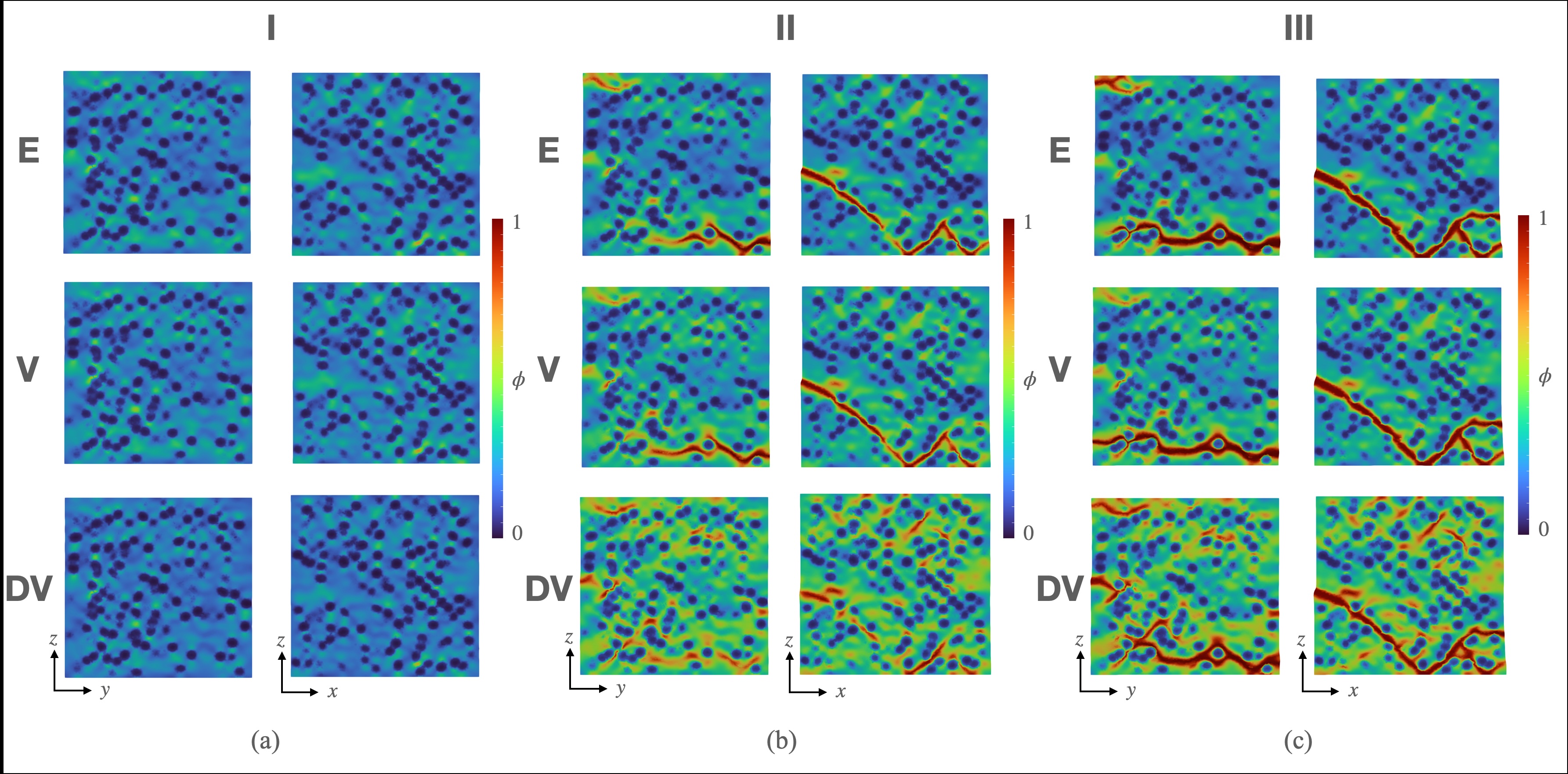}}
      \caption{Granular viscoelastic material deformed in compression, lubricated platens case. Evolution of the phase-field damage variable on $yz$ and $xz$ central cross-sections of the specimen at increasing deformation levels. (a) Peak load (I), (b) 75\% of peak load (II), and (c) 50\% peak load (III). \label{fig_granular_4}}
      \end{figure*}

As expected, increasing the deviatoric viscosity reduces flow stress, whereas increasing the damage viscosity increases the maximum force and produces a more gradual post-peak softening response. Similarly, the model predicts comparable peak forces and applied displacements at full fracture for lubricated and laterally constrained configurations with identical material properties. However, the peak forces are reached at larger applied strains, and the post-peak load decrease is steeper under lubricated conditions. These differences can be explained by examining the corresponding damage evolution and crack propagation patterns in the post-peak midsection slices in Figs. \ref{fig_granular_3} and \ref{fig_granular_4} (see captions for details).

      \begin{figure*}[!h]%
        \centerline{\includegraphics[width=500pt,height=20pc, trim=5 5 8 8,clip]{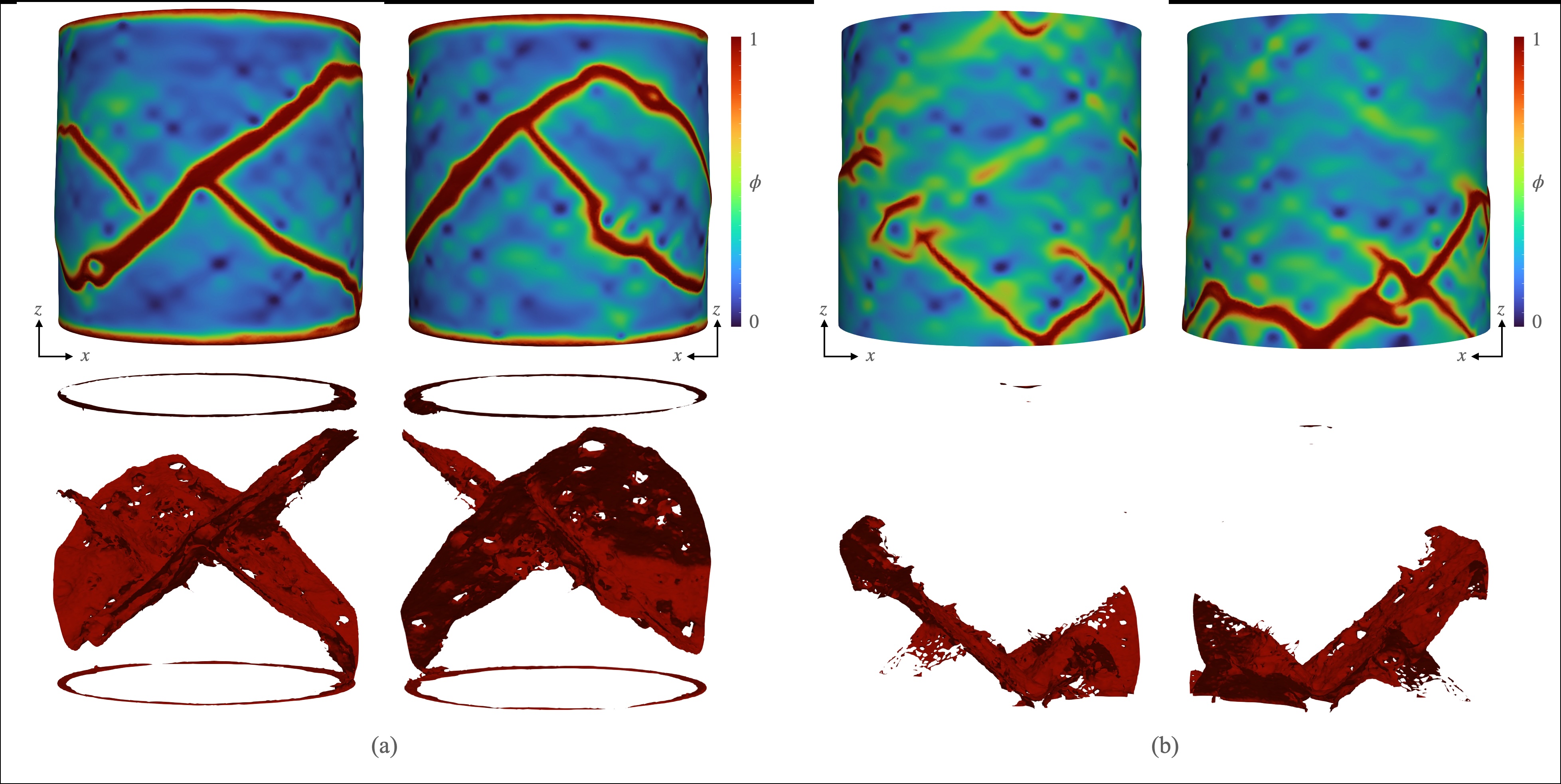}}
        \caption{Granular viscoelastic material deformed in compression. Three-dimensional fracture patterns for (a) laterally constrained, and (b) lubricated platens cases for the composite with a viscoelastic matrix with negligible deviatoric viscosity (V). \label{fig_granular_5}}
        \end{figure*}

When the lateral displacement of the top and bottom surfaces is constrained, dome-shaped "damage-free" regions develop beneath them and damage initiates preferentially at the specimen edges where deviatoric stresses concentrate, as highlighted in Fig. \ref{fig_granular_3}a. Concurrently, damage nucleation sites appear in the central region of the specimen volume and coalesce with increasing applied strain to form a network of long-range cracks, Fig. \ref{fig_granular_3}b. At the final stage of deformation, the central and edge cracks merge, Fig. \ref{fig_granular_3}c. In the three-dimensional renderings of Fig. \ref{fig_granular_5}, this damage evolution results in an X-shaped failure pattern characteristic of brittle granular composites under compression \cite{manner_situ_2017, mehrdad_-situ_2026}. Adding viscoelasticity slows down damage accumulation while maintaining the overall crack pattern largely unchanged, as shown in Fig. \ref{fig_granular_3}c, consistent with damage being primarily driven by deviatoric stresses under compressive loads. By contrast, increasing damage viscosity diffuses damage accumulation, resulting in an increased number of damage nucleation sites at later stages of deformation. This leads to a more pronounced fragmentation of the lateral regions within the X-shaped failure pattern that is qualitatively closer to experimental observations for PBX \cite{manner_situ_2017, mehrdad_-situ_2026}.

Lubricated platens prevent the buildup of deviatoric stresses and associated crack initiation at the specimen edges. As a result, the shadowing effect near the contact surfaces and the accumulation of damage in the central region give way to a more uniform damage distribution, Fig. \ref{fig_granular_4}. Unlike the laterally constrained case, no centrosymmetric X-shaped crack pattern develops. Instead, a dominant crack forms that is confined to the lower half of the specimen, as shown in Fig. \ref{fig_granular_5}b. The resulting crack surface is less extensive than in the X-shaped failure mode, indicating that less fracture energy was expended. The associated collapse of the lower region explains the rapid post-peak load drop predicted for the lubricated case (Fig. \ref{fig_granular_2}b).

\subsection{Particle-reinforced Visco-elastoplastic Material in Tension}

\begin{figure*}[!h]%
  \centerline{\includegraphics[width=500pt,height=17pc, trim=5 5 8 10,clip]{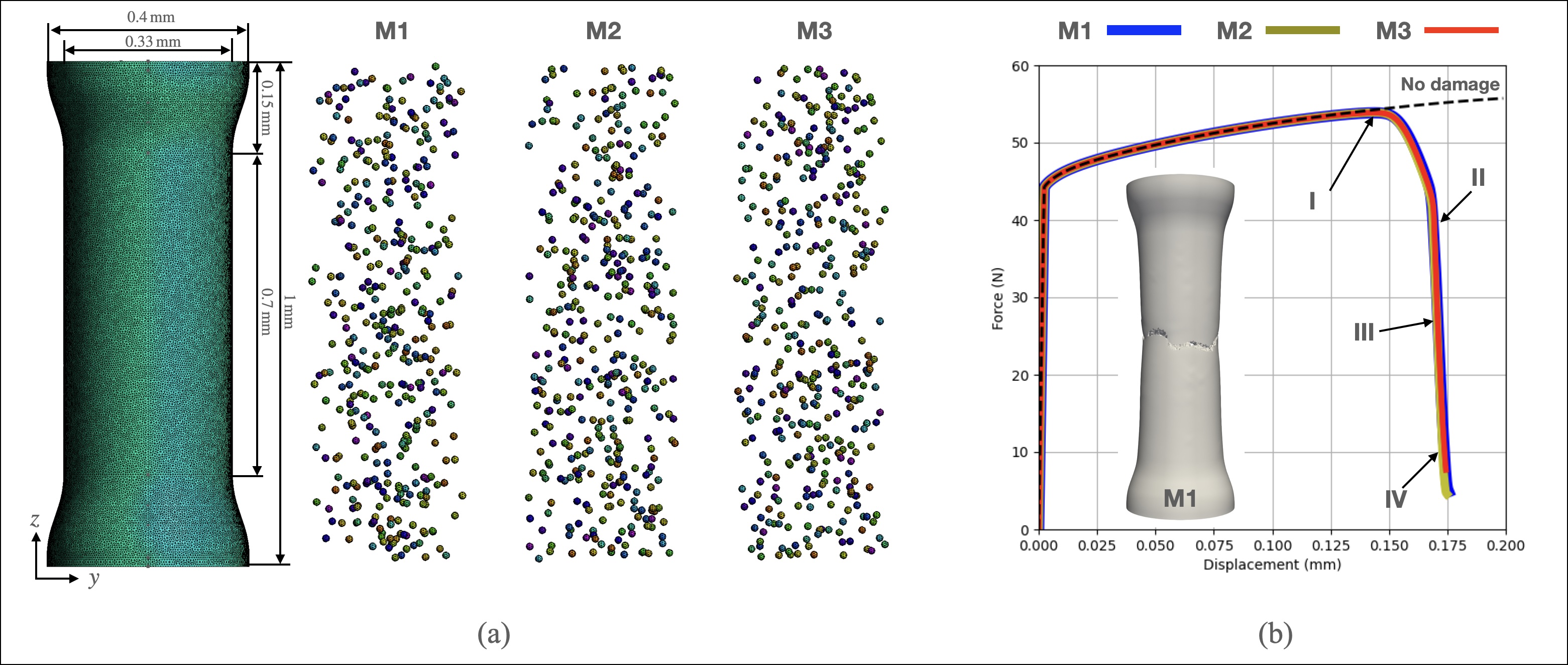}}
  \caption{Particle-reinforced visco-elastoplastic material deformed in tension: Geometry, microstructures, and macroscopic response of the microstructure-resolved tensile specimen. (a) finite element mesh of the dogbone specimen together with the three statistically equivalent particle-matrix microstructures (M1-M3) used in the simulations. (b) Predicted force-displacement curves for the three microstructures. The dashed line indicates the elastic-viscoplastic reference response obtained while suppressing damage. Points I-IV mark peak and post peak stages of the deformation for which accumulated plastic strain and damage fields are reported.\label{fig_alloy_1}}
  \end{figure*}

In this numerical example, the microstructure is varied while the mechanical properties of the two phases are kept fixed. The particle phase consists of about 390 hard spherical $15 \, \mathrm{\mu m}$ inclusions and a mean mesh size of $5 \, \mathrm{\mu m}$. Three different realizations of the microstructure are produced by changing the random number generator in the \texttt{microstructpy} seeding procedure, which varies particle positions while preserving their statistical distributions. Fig. \ref{fig_alloy_1}a shows finite element meshes for the resulting microstructures. The enveloping cylindrical volume has enlarged shoulders to avoid stress concentrations at the laterally constrained ends. The specimen is loaded in uniaxial tension along the cylinder axis at a strain rate of $\approx 4.0 \times 10^{-1} \, \mathrm{s^{-1}}$, measured relative to the gauge length.

\begin{figure*}[!h]%
  \centerline{\includegraphics[width=500pt,height=32pc, trim=5 5 8 5,clip]{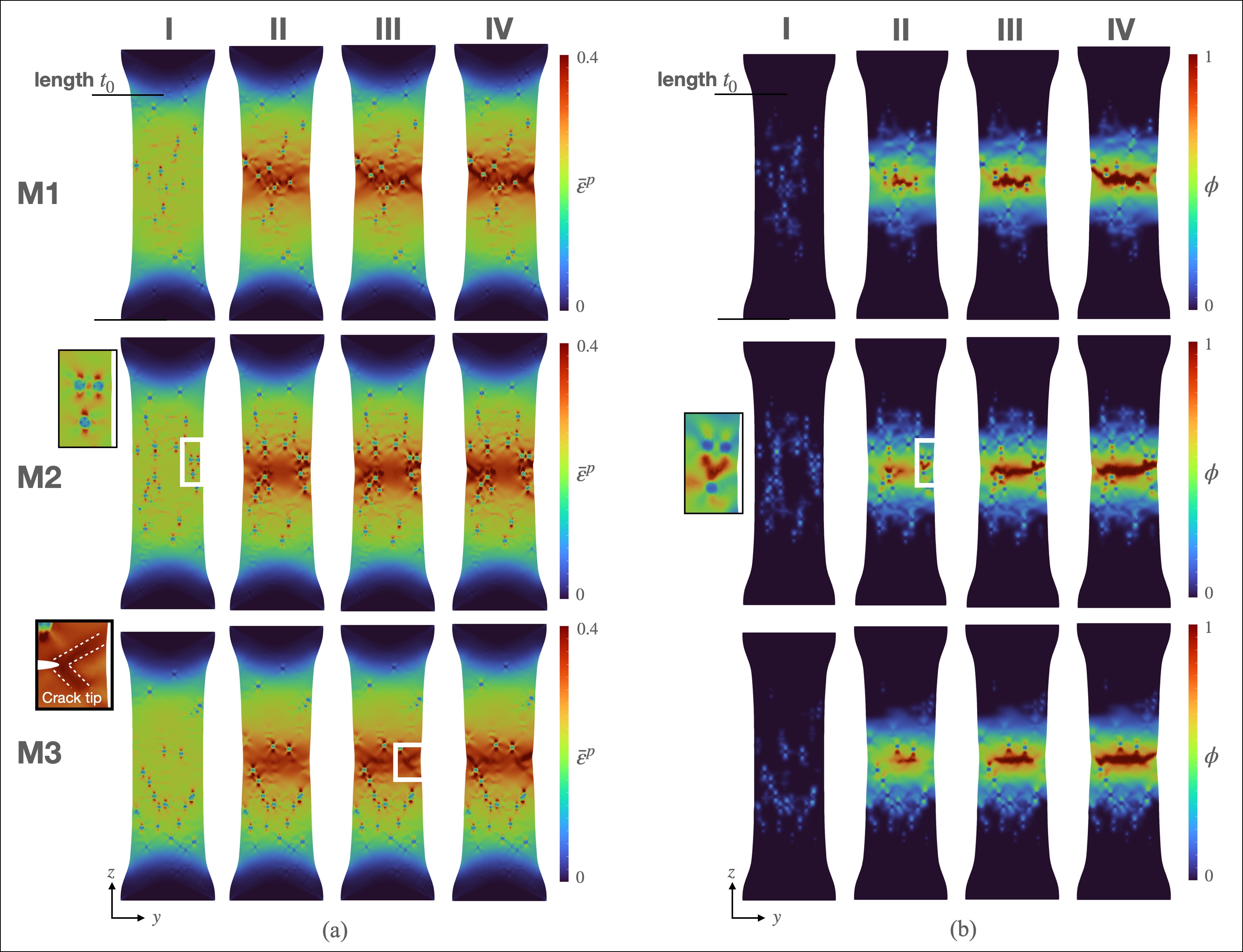}}
  \caption{Particle-reinforced visco-elastoplastic material deformed in tension: Evolution of inelastic deformation and damage during post peak loading for the three statistically equivalent microstructures (M1-M3). (a) Accumulated plastic strain and (b) phase-field damage. Columns correspond to the peak (I) and post peak (II-IV) of applied strain indicated in Fig. \ref{fig_alloy_1}b. \label{fig_alloy_2}}
  \end{figure*}

\begin{figure*}[!h]%
  \centerline{\includegraphics[width=480pt,height=25pc, trim=5 5 8 5,clip]{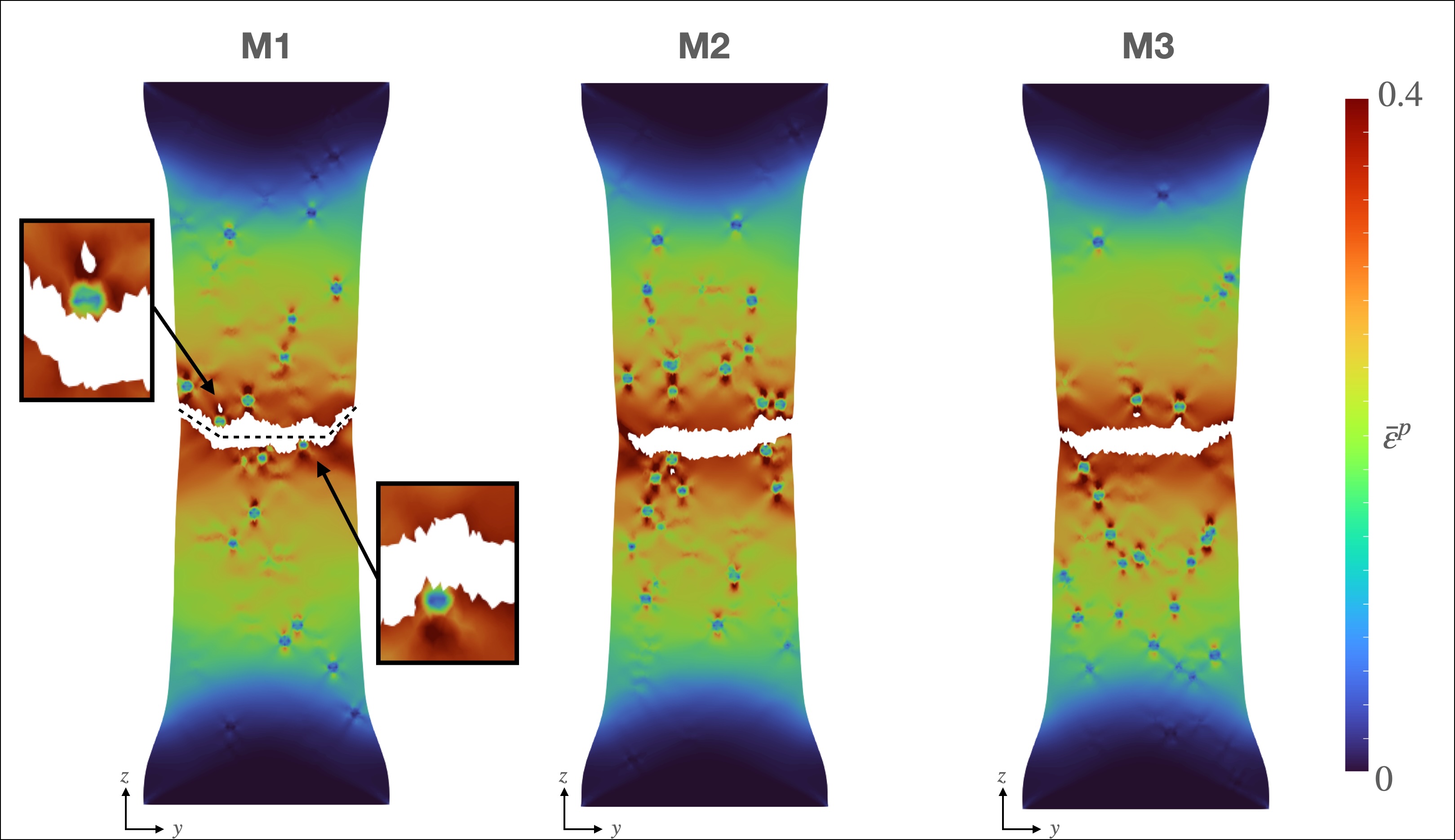}}
  \caption{Particle-reinforced visco-elastoplastic material deformed in tension: Planar cross-sections of accumulated plastic strain predicted at the final applied strain for the three statistically equivalent microstructures (M1-M3). Regions with underlying damage values above 0.95 are masked out. \label{fig_alloy_3}}
  \end{figure*}
  
The predicted force-displacement curves are shown in Fig. \ref{fig_alloy_1}b, together with the reference response obtained for microstructure M1 with $G_c$ high enough to suppress damage over the range of applied strains (dashed line). In this purely viscoplastic case, strain hardening associated with the Hooke and Maxwell branches, together with particle reinforcement, prevents necking and sustains load increase. By contrast, when damage is active, the predicted curves show the characteristic smooth approach to a peak force, followed by gradual post-peak softening. Inspection of the deformed specimens reveals that necking has occurred, as illustrated for M1 in the inset of the same figure. Damage therefore promotes geometric instability, which amplifies the influence of microstructural variability on the post-peak response, as reflected in the slight divergence of the force-displacement curves in this regime.

To further investigate the influence of microstructural variability, Fig. \ref{fig_alloy_2} shows central cross-sections for damage and accumulated plastic strain fields at the I-IV stages of applied deformation indicated on the force-displacement curves. During loading, plastic strain accumulates around the hard particles, forming characteristic deformation plumes (see magnified detail in I-column, Fig. \ref{fig_alloy_2}a). At peak stress, damage localizes in the same regions and preferentially develops within the matrix channels between neighboring particles. Because most deviatoric strain energy ($\psi_\mathrm{d}$) is dissipated through plastic flow, damage is driven primarily by volumetric strain energy contributions, consistent with void nucleation and growth. While damage nucleation promotes necking, the associated local reduction in cross-sectional area is accompanied by stress buildup that, in turn, accelerates damage accumulation. At this stage, voids begin to coalesce within the necking region (see magnified detail in II-column, Fig. \ref{fig_alloy_2}b), giving rise to a disc-shaped crack perpendicular to the loading axis, as commonly observed in experimental observations \cite{tvergaard_analysis_1984}. 

With increasing applied strain, shear bands in the lateral ligaments develop at $ 45^\circ$ to the loading direction, as shown in the magnified detail in column III, Fig. \ref{fig_alloy_2}a. In agreement with experimental observations that have identified shear banding as the precursor of ductile fracture, cracks propagate outward along these bands (column IV in Fig. \ref{fig_alloy_2}b). This behavior corresponds to the well-documented transition from Mode I fracture to shear-dominated Mode II fracture, associated with cup-and-cone fracture patterns observed in conventional uniaxial tensile tests of cylindrical ductile specimens \cite{tvergaard_analysis_1984}. Fig. \ref{fig_alloy_3} better visualizes the characteristic crack pattern, where regions with $\phi > 0.95$ are masked out from accumulated plastic strain maps, and the cup-and-cone profile is highlighted for M1 with dashed lines. Cracks also preferentially nucleate and propagate along the particle-matrix interfaces (see magnified details), exposing particles, as commonly observed in fractographic analysis.

Most relevant to the scope of the present study is the interplay between plastic deformation and damage evolution, which must remain consistent with the rheological assembly shown in Fig. \ref{fig_rheological_model} and, more broadly, with experimental observations. In regions where damage accumulates to the point of full crack formation, further plastic strain accumulation is inhibited, as evidenced by the absence of the disc-shaped localization pattern in the accumulated plastic strain field in Fig. \ref{fig_alloy_2}a. Because damage does not induce plastic softening through yield surface contraction, damage does not promote additional plastic flow where it accumulates. Instead, continued damage evolution leads to rapid loss of load-carrying capacity rather than prolonged necking and gradual softening toward zero load. Experimentally, this manifests itself as an unstable failure characterized by catastrophic rupture and a characteristic snapback response that terminates the tensile test. In the present simulations, this behavior is captured in a quasi-static setting by introducing damage viscosity, which leads to rapid but numerically sustained unloading once crack localization occurs.

\section{Conclusions and Outlook}\label{sec6}

We have presented a phase-field fracture framework for inelastic materials based on Perić \& Dettmer constitutive formulation, designed to enable flexible and efficient modeling of fracture in viscoplastic materials subjected to finite deformations. A rheological \textit{fracture element} was introduced and assembled in series with the Perić \& Dettmer constitutive block, maintaining elastic strain energy as the sole driver of damage while ensuring the latter affects the overall mechanical response without degrading the underlying inelastic properties. This approach has been described as \textit{"brittle fracture in elastic-plastic solids"} in Duda \textit{et al.} \cite{duda2015phase} when applied to homogeneous material volumes. Here, we have shown that the proposed constitutive framework can capture the characteristic shear banding, surface fracture features, and crack patterns that define ductile fracture. This supports our initial hypothesis that the rheological framework is particularly well suited for microstructure-resolved simulations in which damage nucleation and evolution can be identified with void nucleation and microcrack propagation.

We have completed a matrix-free implementation of the proposed constitutive framework using the open source \texttt{Ratel} library. To the best of the authors' knowledge, this represents the first matrix-free phase-field fracture study that includes large strain inelasticity as well as GPU use. In particular, we presented numerical examples executed on the El Capitan high-performance computing prototype, demonstrating \texttt{Ratel} suitability for solving highly nonlinear problems on next-generation GPU architectures and HPC platforms. Owing to the weak coupling between displacement and damage, the use of high-order discretizations, and the regularizing effect of damage viscosity, robust convergence was achieved using a monolithic formulation with $p$-multigrid preconditioning and a four-component solution. In a recent uncertainty quantification study \cite{schmid_calibrating_2025}, \texttt{Ratel} efficient implementation of analogous constitutive models has provided hundreds of high-fidelity numerical simulations to support the development of surrogate models of viscoelastic fracture in mock-PBX material.

Future work will extend the current framework to anisotropic inelasticity through crystal plasticity finite element models (CPFEM). In these models, plastic flow is heterogeneous even in specimens of constant cross-sectional area, as it depends on the Schmid factor of individual grains at the microstructural length scale. This induces a more gradual and realistic transition between elastic and plastic regimes than in the isotropic examples considered here, further justifying the use of perfect plastic Prandtl branch(es). Attention will then focus on studying the role of plastic strain transfer mechanisms, such as slip transmission or the more recently described lattice curvature-mediated mode \cite{di_gioacchino_new_2020}, in preventing damage nucleation at grain boundaries and other microstructural interfaces.
Because lattice curvature and the associated storage of geometrically necessary dislocations (GNDs) are linked to slip gradients \cite{arsenlis_crystallographic_1999}, strain gradient crystal plasticity formulations should be implemented \cite{shu_strain_1999, acharya_lattice_2000}. Numerical simulations of interest would include studying the effect of the resistance to GNDs accumulation on plastic strain transfer and damage across grains. The numerical complexity introduced by high-order plasticity theories, and the loss of coaxiality associated with anisotropic inelasticity, justifies the adoption of automatic differentiation, which has already been successfully applied to simpler material models within \texttt{Ratel}. Unlike the cases considered in the present study, future applications could include hard phases with more complex morphologies and fracture toughness comparable to or lower than that of the surrounding material. Consistent with experimental observations \cite{edwards2017interaction}, numerical simulations should capture crack initiation within these particles and subsequent crack propagation into the surrounding material.

Although the present study demonstrates the capability of the finite element implementation at large strains, CPFEM applications of practical interest involve even more substantial deformations, especially if these can capture slip localization associated with slip bands formation. In these regimes, elastic strains become negligible compared to plastic strains, potentially requiring mixed finite element formulations, which are already available in \texttt{Ratel} for Ogden hyperelasticity. To prevent the loss of numerical convergence associated with the severe distortions of finite elements, we will further aim to extend the proposed framework to the material point method, which has been recently implemented in \texttt{Ratel} and used to simulate the confined compression of PBX materials \cite{atkins_ratel_impm_2026}.

\bmsection*{Author contributions}

\textit{Fabio Di Gioacchino} - Conceptualization and drafting. Constitutive modeling. Implementation (Phase-field fracture, Viscoelasticity, Plasticity, Coupling). Synthetic microstructure, postprocessing, and meshing. Numerical simulations. Data analysis. \textit{Rezgar Shakeri} -  Implementation (Eigenvalues decomposition, Phase-field fracture, Plasticity). \textit{Zachary Atkins} - Implementation (Contact). Numerical simulations. \textit{Karen Stengel} - Implementation (Plasticity). \textit{Layla Ghaffari} - Implementation (Plasticity). \textit{Jeremy Thompson} - Implementation (Ratel infrastructure and matrix-free methods). \textit{Jed Brown} - Implementation (Matrix-free methods). Funding procurement. -- All authors have contributed to the manuscript.

\bmsection*{Acknowledgments}
This work was supported by the Department of Energy, National Nuclear Security Administration under the Predictive Science Academic Alliance Program (PSAAP), Award Number DE-NA0003962. The authors also acknowledge support by the Department of Energy Frameworks, Algorithms, and Scalable Technologies for Mathematics (FASTMath) SciDAC Institute. The authors thank Prof. Christian Linder and Prof. Lampros Svolos for helpful feedback and discussions on phase-field modeling of ductile fracture.

\bmsection*{Financial disclosure}

None reported.

\bmsection*{Conflict of interest}

The authors declare no potential conflict of interests.

\bibliography{paper}



\appendix

\bmsection{principal stresses-based format}

Once the elastic trial left Cauchy–Green tensor $\bm{b}^{e \,\mathrm{tr}}$ is obtained (see Eq. \eqref{eq:inelasticity-be-update}), it is spectrally decomposed such that
\begin{equation}
  \bm{b}^{e \, \mathrm{tr}} = \sum_{i=1}^3 b^{e\, \mathrm{tr}}_i \,
  \hat{\bm{n}}_i \otimes \hat{\bm{n}}_i,
  \label{eq:inelasticity-spectral-decomposition}
\end{equation}
where $b^{e \, \mathrm{tr}}_i$ and $\hat{\bm{n}}_i$ denote its sets of eigenvalues and eigenvectors, respectively. The eigenvalues of the logarithmic elastic strain are then computed as
\begin{equation}
  \varepsilon_i^{e \, \mathrm{tr}} = \frac{1}{2}\log(b_i^{e \, \mathrm{tr}}),
  \label{eq:inelasticity-log-strain-principal}
\end{equation}

The deviatoric part of the elastic strain is updated independently in each principal direction. For the elastic, viscoelastic and plastic branches considered in this work, the update takes the form
\begin{equation}
  \varepsilon_{\mathrm{d} \, i}^e =
  \begin{cases}
    
    \varepsilon_{\mathrm{d} \, i}^{e \, \mathrm{tr}}
    & \mathrm{HOOKE} \\[6pt]
    \left(1 + \mu\frac{\Delta t}{\eta_\mathrm{d}}\right)^{-1}
    \varepsilon_{\mathrm{d} \, i}^{e \, \mathrm{tr}}
    & \mathrm{MAXWELL} \\[6pt]
    \left(1 - 3\mu\frac{\Delta\gamma}{q^{\mathrm{tr}}}\right)
    \varepsilon_{\mathrm{d}\,i}^{e \, \mathrm{tr}}
    & \mathrm{PRANDTL}
  \end{cases}
  \label{eq:principal-directions-format-2}
\end{equation}
The updated eigenvalues for $\tau$ and $\bm{b}^e$ are then given by
\begin{equation}
  \tau_i = 2\mu\, \varepsilon_{\mathrm{d} \, i}^e + \kappa \mathrm{tr}(\bm{\varepsilon}^e) 
  \qquad \mathrm{with} \qquad
  \mathrm{tr}(\bm{\varepsilon}^e) = \sum_{i=1}^3 \varepsilon_i^e
  \label{eq:principal-stress-update-eigenvalues}
\end{equation}
and
\begin{equation}
  b_i^e = \exp(2 \varepsilon_i^e) 
  \qquad \mathrm{with} \qquad  
  \varepsilon_i^e
  =
  \varepsilon_{\mathrm{d} \, i}^e
  +
  \frac{1}{3}\mathrm{tr}(\bm{\varepsilon}^{e}), 
  \label{eq:be-update-eigenvalues}
\end{equation}
The respective full tensors (stored in Voigt representation) are finally reconstructed from the outer product of common eigenvectors, \textit{i.e.},
\begin{equation}
  \bm{\tau}
  = \sum_{i=1}^{3} \tau_i  \hat{\bm{n}}_i \otimes \hat{\bm{n}}_i
  \label{eq:principal-stress-update-eigenvalues-2}
\end{equation}
and
\begin{equation}
  \bm{b}^e
  = \sum_{i=1}^{3} b^e_i  \hat{\bm{n}}_i \otimes \hat{\bm{n}}_i
  \label{eq:be-update-eigenvalues-2}
\end{equation}

The principal directions format can be extended to $\psi^+$ in Eq. \eqref{eq:positive_branch_energy}:
\begin{equation}
  \psi_k^+
  =
  \frac{\kappa_k}{2}
  \left\langle
    \mathrm{tr}
    (\bm\varepsilon_k^e)
  \right\rangle_+^2
  +
  \mu_k\sum_{i=1}^{3}(\varepsilon_{\mathrm{d}\,i}^{e})^2
  \label{eq:positive_branch_energy-eigen}
\end{equation}

\bmsection{Peric \& Dettmer block consistent tangent}

libCEED allows the consistent linearization of constitutive models through variations of relevant quantities directly at the quadrature level. This approach avoids the explicit assembly and storage of fourth-order tangent tensors in Eq. \eqref{eq:inelasticity-consistent-tangent}, significantly reducing memory usage and enabling efficient matrix-free implementations.

Starting from the algorithmic expression of the left Cauchy-Green tensor in Eq. \eqref{eq:inelasticity-be-update}, its variation with respect to $\bm{F}_{n+1}$, which corresponds to the last term on the right-hand side of Eq. \eqref{eq:inelasticity-consistent-tangent}, is obtained in full representation as
\begin{equation}
  d\bm{b}^{e \, \mathrm{tr}}_{n+1}
  =
  d\bm{F}_{n+1}
  \bm{C}^{in-1}_{n}
  \bm{F}_{n+1}^{T}
  +
  \bm{F}_{n+1}
  \bm{C}^{in-1}_{n}
  d\bm{F}_{n+1}^{T}
\end{equation}

Using its eigenvalue decomposition, the remaining terms can be efficiently constructed from the principal strain derivatives
\begin{equation}
  d \varepsilon^{e \, \mathrm{tr}}_{i}
  = \frac{1}{2 b^{e \, \mathrm{tr}}_i} d b^{e \, \mathrm{tr}}_i
\end{equation}
with 
\begin{equation}
  d \varepsilon^{e \, \mathrm{tr}}_{\mathrm{d} \, i}
  =
  d \varepsilon^{e \, \mathrm{tr}}_{i}
  - \frac{1}{3}\mathrm{tr}(d\bm{\varepsilon}^{e \, \mathrm{tr}}),
  \qquad
  \mathrm{with}
  \qquad
  \mathrm{tr}(d\bm{\varepsilon}^{e \, \mathrm{tr}})  = \sum_{i=1}^{3} d \varepsilon^{e \, \mathrm{tr}}_{i}
\end{equation}

The updated counterpart $ d \varepsilon^{e}_{\mathrm{d} \, i}$ depends on the specific rheological element, see BOX A1. The stress derivative in the principal directions then follows directly as
\begin{equation}
  d\tau_i = 2 \mu  d\varepsilon^e_{\mathrm{d} \, i} + \kappa  \mathrm{tr}(d\bm{\varepsilon}^e)
\end{equation}
Finally, the corresponding isotropic tensor-valued function is assembled as
\begin{equation}
  d\bm{\tau}
  = \sum_{i=1}^{3}
  \left(
    d\tau_i \,
    \hat{\bm{n}}_i \otimes \hat{\bm{n}}_i
    + \tau_i
    \left(
      d\hat{\bm{n}}_i \otimes \hat{\bm{n}}_i
      + \hat{\bm{n}}_i \otimes d\hat{\bm{n}}_i
    \right)
  \right)
  \label{eq:hencky-stress-linearization}
\end{equation}
where the eigenvectors linearization (shown here only for the general case of distinct eigenvalues) is given by
\begin{equation}
  d\hat{\bm{n}}_i
  = \sum_{j \neq i}
    \frac{\hat{\bm{n}}_j(\hat{\bm{n}}_j \cdot d\bm{b}^{e \,\mathrm{tr}}
        \cdot \hat{\bm{n}}_i)
    }{b^{e \, \mathrm{tr}}_i - b^{e \, \mathrm{tr}}_j} 
\end{equation}

\begin{boxwithhead}
  {BOX A1\quad Linearized deviatoric stress updates in principal directions}
  {\noindent 

      \textbf{Hooke element:}
\begin{equation}
  d \varepsilon^{e}_{\mathrm{d} \,i} = d \varepsilon^{e\,\mathrm{tr}}_{\mathrm{d} \,i}
  \label{eq:hencky-dtau-hooke}
\end{equation}

    \textbf{Maxwell branch:}
\begin{equation}
  d \varepsilon^{e}_{\mathrm{d} \,i}
  = \left(1 + \mu\frac{\Delta t}{\eta_\mathrm{d}}\right)^{-1}
    d \varepsilon^{e\,\mathrm{tr}}_{\mathrm{d} \,i}
  \label{eq:hencky-dtau-maxwell}
\end{equation}

    \textbf{Prandtl branch:}

For the Prandtl branch with von Mises plasticity, the linearization of the deviatoric elastic strain follows from differentiating Eq. \eqref{eq:plasticity-deviatoric-strain-update} and using Eq. \eqref{eq:plasticity-delta-gamma-update},
\begin{equation}
\begin{aligned}
  d \varepsilon^{e}_{\mathrm{d} \,i}
  &=
  d\left[
    \left(
      1 - 3\mu \frac{\Delta\gamma}{q^{\mathrm{tr}}}
    \right)
    \varepsilon^{e\,\mathrm{tr}}_{\mathrm{d} \,i}
  \right] \\
  &=
  \left(
    1 - 3\mu \frac{\Delta\gamma}{q^{\mathrm{tr}}}
  \right)
  d \varepsilon^{e\,\mathrm{tr}}_{\mathrm{d} \,i}
  -
  3\mu
  \frac{dq^{\mathrm{tr}}}{q^{\mathrm{tr}}}
  \left(
    \frac{1}{3\mu + H}
    - \frac{\Delta\gamma}{q^{\mathrm{tr}}}
  \right)
  \varepsilon^{e\,\mathrm{tr}}_{\mathrm{d} \,i}
\end{aligned}
\label{eq:plasticity-dtau-vonMises}
\end{equation}
with
\begin{equation}
  d q^{\mathrm{tr}}
  =
  \tfrac{3}{2}
  \frac{
    \bm{\tau}^{\mathrm{tr}}_\mathrm{d} :
    d\bm{\tau}^{\mathrm{tr}}_\mathrm{d}
  }{
    q^{\mathrm{tr}}
  }
  \label{eq:plasticity-dq-trial-vonMises}
\end{equation}
  }
\end{boxwithhead}

\bmsection{Consistent tangent for the monolithic scheme}

The consistent linearization of the degraded Kirchhoff stress in Eq. \eqref{eq:total_degraded_stress} must account for both the explicit dependence of the stress on the phase field through the degradation function and its implicit dependence through the evolution of the inelastic internal variables. Its total variation is
\begin{equation}
  \delta\bm{\tau}_{\mathrm{degr}}
  =
  \delta_{\bm u}\bm{\tau}_{\mathrm{degr}}
  +
  \delta_\phi\bm{\tau}_{\mathrm{degr}}
\end{equation}
with
\begin{equation}
  \delta\bm{\tau}_{\mathrm{degr}}
  =
  \sum_k\Big[
    g(\phi)\delta_{\bm u}\bm{\tau}_k^+
    +\delta_{\bm u}\bm{\tau}_k^-
    +g'(\phi)\bm{\tau}_k^+\delta\phi
    +g(\phi)\delta_\phi\bm{\tau}_k^+
    +\delta_\phi\bm{\tau}_k^-
  \Big]
  \label{eq:linear-degraded-stress-linearization}
\end{equation}

For the isochoric viscous and plastic evolution considered here, the implicit dependence of the branch stresses on the phase field is confined to their deviatoric components. Thus,
$\delta_\phi\bm{\tau}_k^-=0$ and
$\delta_\phi\bm{\tau}_k^+
=\delta_\phi\bm{\tau}_{d,k}$, giving
\begin{equation}
  \delta\bm{\tau}_{\mathrm{degr}}
  =
  \sum_k\Big[
    g(\phi)\delta_{\bm u}\bm{\tau}_k^+
    +\delta_{\bm u}\bm{\tau}_k^-
    +g'(\phi)\bm{\tau}_k^+\delta\phi
    +g(\phi)\delta_\phi\bm{\tau}_{\mathrm{d},k}
  \Big]
  \label{eq:linear-degraded-stress-isochoric}
\end{equation}
Here,
\begin{equation}
  \delta_\phi\bm{\tau}_{\mathrm{d},k}
  :=
  \frac{\partial\bm{\tau}_{\mathrm{d},k}}
       {\partial\phi}\delta\phi
\end{equation}
where $\bm{\tau}_{\mathrm{d},k}$ denotes the undegraded deviatoric spring stress evaluated at the updated elastic strain, and $g'(\phi)=-2(1-\phi)$.

For the viscoelastic Maxwell branch,
\begin{equation}
  \frac{\partial\bm{\tau}_{\mathrm{d}}}{\partial\phi}
  =
  -2\mu\frac{\mu\Delta t}{\eta_{\mathrm{d}}}
  \frac{g'(\phi)}
  {\left(1+\dfrac{\mu\Delta t}{\eta_{\mathrm{d}}}g(\phi)\right)^2}
  \bm{\varepsilon}_{\mathrm{d}}^{e\,\mathrm{tr}}
  =
  -\frac{\mu\Delta t}{\eta_{\mathrm{d}}}
  \frac{g'(\phi)}
  {1+\dfrac{\mu\Delta t}{\eta_{\mathrm{d}}}g(\phi)}
  \bm{\tau}_{\mathrm{d}}
  \label{eq:maxwell-dtau-dphi}
\end{equation}

For the elastoplastic Prandtl branch in the perfect plastic regime,
\begin{equation}
  \frac{\partial\bm{\tau}_{\mathrm{d}}}{\partial\phi}
  =
  -2\mu\frac{\sigma_0}{q^{\mathrm{tr}}}
  \frac{g'(\phi)}{g(\phi)^2}
  \bm{\varepsilon}_{\mathrm{d}}^{e\,\mathrm{tr}}
  =
  -\frac{g'(\phi)}{g(\phi)}
  \bm{\tau}_{\mathrm{d}}
  \label{eq:prandtl-dtau-dphi}
\end{equation}
Consistent with the result in Eq. \eqref{eq:degraded-perfect-plastic-stress}, Eq. \eqref{eq:prandtl-dtau-dphi} makes the contribution to the degraded deviatoric stress independent of $\phi$,
\begin{equation}
  \frac{\partial\bm{\tau}_{\mathrm{d \, degr}}}{\partial\phi}
  =
  g'(\phi)\bm{\tau}_{\mathrm{d}}
  +
  g(\phi)\frac{\partial\bm{\tau}_{\mathrm{d}}}{\partial\phi}
  =
  \bm{0}
  \label{eq:prandtl-degraded-dtau-dphi}
\end{equation}

The linearization of $L$ in Eq. \eqref{eq:L-viscous} reads
\begin{equation}
  \delta L
  =
  \left[
    g''(\phi)\mathcal{H}
    +
    \frac{G_c}{c_0l_0}\alpha''(\phi)
    +
    \frac{\zeta}{\Delta t}
  \right]\delta\phi
  +
  g'(\phi)\delta\mathcal{H}
  \label{eq:linearized-local-damage-residual}
\end{equation}
with $g''(\phi)=2$, $\alpha''(\phi)=0$ for AT1, and $\alpha''(\phi)=2$ for AT2.
The variation $\delta\mathcal{H}$ is subject to the irreversibility condition
\begin{equation}
  \delta\mathcal{H}_{n+1}
  =
  \begin{cases}
    \delta\psi_{n+1}^{+}=
  \delta_{\bm{u}}\psi_{n+1}^{+}
  +
  \delta_{\phi}\psi_{n+1}^{+}
    & \psi_{n+1}^{+}>\mathcal{H}_n
    \\[1ex]
    0
    & \psi_{n+1}^{+}<\mathcal{H}_n
  \end{cases}
  \label{eq:history-field-linearization}
\end{equation}
For all branches,
\begin{equation}
  \delta_{\bm{u}}\psi^{+}
  =
  \begin{cases}
    \bm{\tau}:\delta_{\bm{u}}\bm{\varepsilon}^{e},
    & \mathrm{tr}(\bm{\varepsilon}^e)\geq0
    \\[6pt]
    \bm{\tau}_{\mathrm{d}}:
    \delta_{\bm{u}}\bm{\varepsilon}_{\mathrm{d}}^{e},
    & \mathrm{tr}(\bm{\varepsilon}^e)<0
  \end{cases}
  \label{eq:energy-displacement-linearization}
\end{equation}
For the purely elastic Hooke branch, $\psi^{+}$ is independent of $\phi$, and therefore
\begin{equation}
  \delta_{\phi}\psi^{+}=0.
\end{equation}
For the Maxwell branch,
\begin{equation}
\begin{aligned}
  \delta_{\phi}\psi^{+}
  =
  \delta_{\phi}\psi_{\mathrm{d}}
  &=
  \bm{\tau}_{\mathrm{d}}:
  \frac{\partial\bm{\varepsilon}_{\mathrm{d}}^{e}}
       {\partial\phi}\,\delta\phi
  \\
  &=
  -\bm{\tau}_{\mathrm{d}}:
  \frac{\mu\Delta t}{\eta_{\mathrm{d}}}
  \frac{g'(\phi)}
  {\left(1+\dfrac{\mu\Delta t}{\eta_{\mathrm{d}}}g(\phi)\right)^2}
  \bm{\varepsilon}_{\mathrm{d}}^{e\,\mathrm{tr}}\delta\phi
  \\
  &=
  -\bm{\tau}_{\mathrm{d}}:
  \frac{\mu\Delta t}{\eta_{\mathrm{d}}}
  \frac{g'(\phi)}
  {1+\dfrac{\mu\Delta t}{\eta_{\mathrm{d}}}g(\phi)}
  \bm{\varepsilon}_{\mathrm{d}}^{e}\delta\phi
  \\
  &=
  -2\frac{\mu\Delta t}{\eta_{\mathrm{d}}}
  \frac{g'(\phi)}
  {1+\dfrac{\mu\Delta t}{\eta_{\mathrm{d}}}g(\phi)}
  \psi_{\mathrm{d}}\delta\phi
\end{aligned}
\label{eq:maxwell-energy-dphi}
\end{equation}
For the Prandtl branch in the perfect plastic regime,
\begin{equation}
\begin{aligned}
  \delta_{\phi}\psi^{+}
  =
  \delta_{\phi}\psi_{\mathrm{d}}
  &=
  \bm{\tau}_{\mathrm{d}}:
  \frac{\partial\bm{\varepsilon}_{\mathrm{d}}^{e}}
       {\partial\phi}\,\delta\phi
  \\
  &=
  \bm{\tau}_{\mathrm{d}}:
  \left[
    -\frac{g'(\phi)\sigma_0}
    {g(\phi)^2q^{\mathrm{tr}}}
    \bm{\varepsilon}_{\mathrm{d}}^{e\,\mathrm{tr}}
  \right]\delta\phi
  \\
  &=
  -\frac{g'(\phi)}{g(\phi)}
  \bm{\tau}_{\mathrm{d}}:
  \bm{\varepsilon}_{\mathrm{d}}^{e}\delta\phi
  \\
  &=
  -2\frac{g'(\phi)}{g(\phi)}
  \psi_{\mathrm{d}}\delta\phi
\end{aligned}
\label{eq:prandtl-energy-dphi}
\end{equation}




\end{document}